\documentclass[12pt]{article}
\usepackage{scicite}
\usepackage{times}
\usepackage{amsmath}
\usepackage{amsfonts}
\usepackage{amssymb}
\usepackage{graphicx}
\usepackage{subfigure}
\usepackage{bm}
\usepackage{color}
\usepackage[colorlinks=true,citecolor=blue,urlcolor=blue,linkcolor=blue,hyperfigures=true]{hyperref}
\usepackage{multirow}
\usepackage{comment}
\usepackage{physics}

\def\Eq{Eq.~}
\def\Eqs{Eqs.~}
\def\Fig{Fig.~}
\def\Figs{Figs.~}
\def\Ref{Ref.~}

\def\be{\begin{equation}}
\def\ee{\end{equation}}
\def\ba{\begin{align}}
\def\ea{\end{align}}
\def\ie{\textit{i.e.}~}
\def\eg{\textit{e.g.}~}

\def\dd{\mathrm{d}}

\newenvironment{sciabstract}{%
\begin{quote} \bf}
{\end{quote}}

\title{Tracking the Vector Acceleration with a Hybrid Quantum Accelerometer Triad}

\author
{Simon Templier,$^{1,2}$
Pierrick Cheiney,$^{1}$
Quentin d'Armagnac de Castanet,${}^{1,2}$\\
Baptiste Gouraud,$^{1}$
Henri Porte,$^{1}$
Fabien Napolitano,$^{1}$
Philippe Bouyer,$^{1,3,4,5}$\\
Baptiste Battelier,$^{2,\ast}$
and Brynle Barrett$^{1,2,5}$
\\
\normalsize{${}^{1}$iXblue, 34 rue de la Croix de Fer, 78105 Saint-Germain-en-Laye, France}\\
\normalsize{${}^{2}$LP2N, Laboratoire Photonique, Num\'{e}rique et Nanosciences,}\\
\normalsize{Universit\'{e} Bordeaux--IOGS--CNRS:UMR 5298,}\\
\normalsize{1 rue Fran\c{c}ois Mitterrand, 33400 Talence, France}\\
\normalsize{${}^{3}$Van der Waals-Zeeman Institute, Institute of Physics,}\\
\normalsize{University of Amsterdam, Science Park 904, 1098XH Amsterdam, The Netherlands}\\
\normalsize{${}^{4}$QuSoft, Science Park 123, 1098XG Amsterdam, The Netherlands}\\
\normalsize{${}^{5}$Eindhoven University of Technology, The Netherlands}\\
\normalsize{${}^{5}$Department of Physics, University of New Brunswick,}\\
\normalsize{8 Bailey Dr., Fredericton NB, E3B 5A3, Canada}
\\
\normalsize{$^\ast$To whom correspondence should be addressed; E-mail:  baptiste.battelier@institutoptique.fr.}
}

\date{\today}

\begin{document}
%%%%%%%%%%%%%%%%%%%%%%%%%%%%%%%%%%%%%%%%%%%%%%%%%%%%%%%%%%%%%%%%%%%%%%%%%%%%%%%%%%%%%%%%%%%%%%%%%


\baselineskip24pt


\maketitle

%================================================================================================


% The abstract should be a single paragraph, not to exceed 250 words and ideally closer to 200, written in plain language that a general reader can understand. It should include
% An opening sentence that states the question/problem addressed by the research AND
% Enough background content to give context to the study AND
% A brief statement of primary results AND
% A short concluding sentence.
% Do not include citations or undefined abbreviations in the abstract. Any abbreviations that appear in the title should be defined in the abstract.

\begin{sciabstract}
Robust and accurate acceleration tracking remains a challenge in many fields. For geophysics and economic geology, precise gravity mapping requires onboard sensors combined with accurate positioning and navigation systems. Cold-atom-based quantum inertial sensors can potentially provide such high-precision instruments. However, current scalar instruments require precise alignment with vector quantities. Here, we present the first hybrid three-axis accelerometer exploiting the quantum advantage to measure the full acceleration vector by combining three orthogonal atom interferometer measurements with a classical navigation-grade accelerometer triad. Its ultra-low bias permits tracking the acceleration vector over long timescales---yielding a 50-fold improvement in stability ($6 \times 10^{-8}~g$) over our classical accelerometers. We record the acceleration vector at a high data rate (1 kHz), with absolute magnitude accuracy below 10 $\mu g$, and pointing accuracy of 4 $\mu$rad. This paves the way toward future strapdown applications with quantum sensors and highlights their potential as future inertial navigation units.
\end{sciabstract}

% In setting up this template for *Science Advances* papers, both
% the \section* command and the \paragraph* command are used for topical
% divisions.  Which you use will of course depend on the type of paper
% you're writing.  Review Articles tend to have displayed headings, for
% which \section* is more appropriate; Research Articles, when they have
% formal topical divisions at all, tend to signal them with bold text
% that runs into the paragraph, for which \paragraph* is the right
% choice.  Either way, use the asterisk (*) modifier, as shown, to

%================================================================================================
\section*{Introduction}
\label{sec:Intro}
% The manuscript should start with a brief introduction that lays out the problem addressed by the research and describes the paper's importance. The scientific question being investigated should be described in detail. The introduction should provide sufficient background information to make the article understandable to readers in other disciplines, and provide enough context to ensure that the implications of the experimental findings are clear.

Our ability to manipulate and control light and matter at the quantum level has opened a suite of {\em quantum technologies} that promises to provide revolutionary new sensors that feature both high accuracy and high sensitivity for a large variety of applications. Their ability to measure minute changes in inertial quantities such as accelerations and rotations with unprecedented performance in terms of accuracy, sensitivity, and long-term stability can lead to paradigmatic changes in our ability to navigate without external aid \cite{Jekeli2005}, monitor our planet \cite{Leveque2021}, or test the predictions of physical theories \cite{Rosi2014, Parker2018, Morel2020, Asenbaum2020, Zhou2021, Barrett2022}. Today, matter-wave inertial sensors \cite{Borde1989, Kasevich1991} provide mature accelerometers that can measure gravity \cite{Freier2016, Hardman2016, Menoret2018, Bidel2020} and gravity gradients \cite{Xu2017, Savoie2018, Moan2019, Caldani2019, Janvier2022} to remotely detect massive objects or mass movements, and can be used to anticipate major risks such as earthquakes, volcanic eruptions, and sea-level rise \cite{Bongs2019}. These quantum accelerometers are also key elements for future autonomous positioning and navigation devices.

Acceleration is a vector and is therefore described by both its absolute magnitude (or norm) and pointing direction. The majority of matter-wave inertial sensors realized so far are scalar in nature---they measure the projection of the acceleration on a preferred orientation defined by the interrogation laser used for quantum manipulation. Laboratory-based gravimeters thus rely on a precise orientation with respect to the vertical direction \cite{Freier2016, Menoret2018}, while mobile operation requires mounting the sensor on a gyro-stabilized platform \cite{Bidel2018, Bidel2020}. In both cases, real-time tracking of the acceleration is complex and thus limits the potential application of these sensors until a fully three-dimensional (3D) vector-type sensor \cite{Barrett2019} is available. Several groups have made encouraging progress with multi-axis sensing architectures \cite{Canuel2006, Dickerson2013, Wu2017, Chen2019, Gautier2022}. However, none of these studies have yet demonstrated a robust, motion compatible instrument capable of measuring the full acceleration vector in real time.

%--------------------------------------------------
\begin{figure}[!t]
  \centering
  \includegraphics[width=\textwidth]{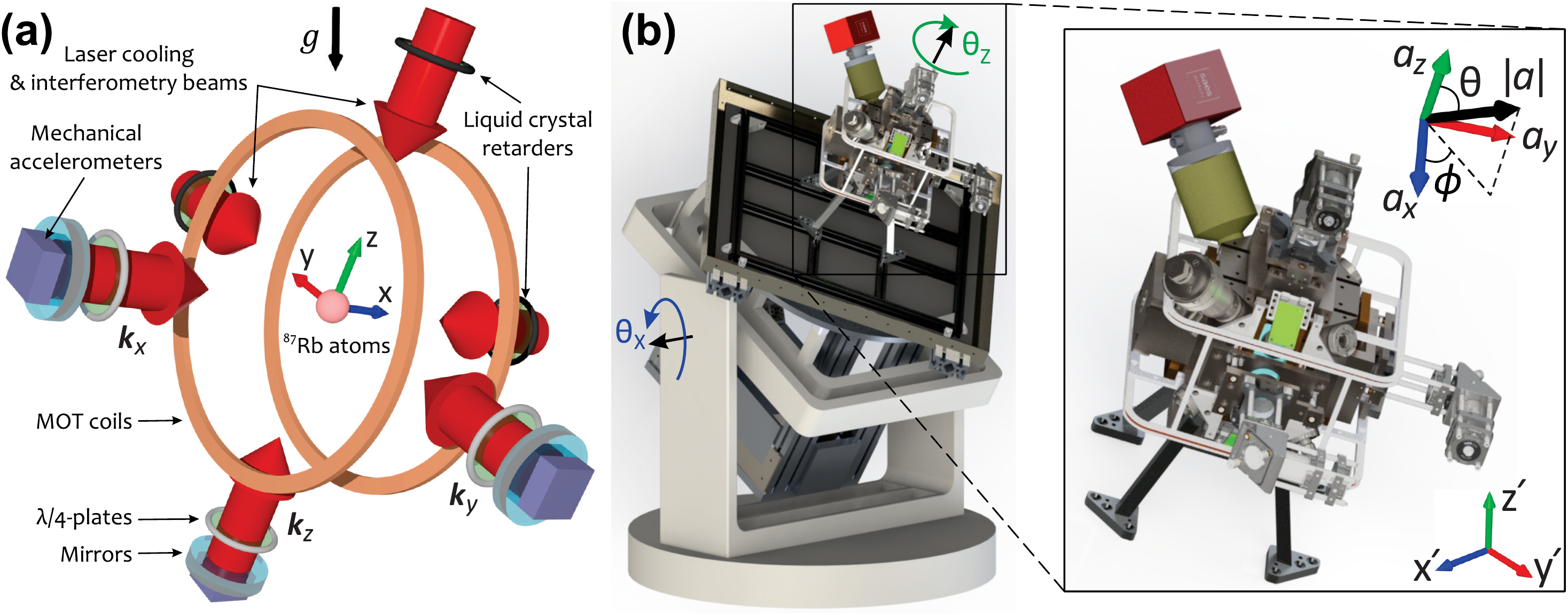}
  \caption{(a) Design concept and geometry of the Quantum Accelerometer Triad (QuAT). The acceleration components are measured along the wavevectors $\bm{k}_x$, $\bm{k}_y$, and $\bm{k}_z$, which are perpendicular to the surface of their respective mirrors. (b) 3D model of the sensor head mounted on a rotation stage that can be tilted about the $z$- and $x'$-axes by angles $\theta_z$ and $\theta_x$, respectively. Applied in this order, these extrinsic rotations transform the coordinates from the laboratory frame ($x'y'z'$) to the body frame ($xyz$) of the QuAT. A cube-shaped $\mu$-metal magnetic shield (not shown) surrounds the entire sensor head. The inset shows the Cartesian ($a_x, a_y, a_z$) and spherical polar (norm $|a|$, inclination $\theta$, azimuth $\phi$) coordinate representations of the acceleration vector in the body frame.}
  \label{fig:3DAccelDesign}
\end{figure}
%--------------------------------------------------

Here, we report the first Quantum Accelerometer Triad (QuAT), which measures accelerations along three mutually-orthogonal directions. Our QuAT, shown in \Fig \ref{fig:3DAccelDesign}, consist of a hybrid 3D architecture combining cold atoms and classical accelerometers \cite{Merlet2009, Barrett2016a, Cheiney2018} to achieve high data rate (1 kHz), ultra-low bias ($\sim 5$ $\mu g$) measurements of the three components of acceleration ($a_x$, $a_y$, $a_z$), with excellent long-term stability---reaching 60 n$g$ on the vector norm ($|\bm{a}| = \sqrt{a_x^2 + a_y^2 + a_z^2}$) after 24 h of integration in static conditions. After adapting broadly-used accelerometer calibration methods to our hybrid triad \cite{Yang2012}, we estimate an accuracy of 7.7 $\mu g$ on the vector norm, which is limited primarily by small misalignments between axes ($\sim 4$ $\mu$rad). Further improvements to the mechanical constraints of the system are anticipated to reduce systematics and improve accuracy by an order of magnitude. At this level, tilts of the tidal gravitational anomaly could be measured and used, for instance, to correct effects on large-area optical gyroscopes \cite{Schreiber2003}, or to further improve models of Earth's gravity \cite{Timmen1995, Xu2017}. This work also paves the way toward strapdown inertial navigation with quantum sensors, and opens new possibilities for gravity mapping, mineral exploration, seismology, and monitoring climate change.

%================================================================================================
\section*{Results}
\label{sec:Results}

% The results should describe the experiments performed and the findings observed. The results section should be divided into subsections to delineate different experimental themes. All data must be shown either in the main text or in the Supplementary Materials.
% All data should be presented in the Results. No data should be presented for the first time in the Discussion.
% All data must be shown; references to "unpublished results" or "data not shown" are not permitted.

Each axis of the QuAT consists of a laser beam retro-reflected by a mirror. These three orthogonal mirrors define the 3D reference frame with respect to which we measure the atom's motion (one axis at a time). Classical accelerometers are attached to the rear of each mirror to monitor their motion, as well as the component of gravitational acceleration. We then correct the frequency and phase of each laser to account for the motion the atom relative to the corresponding mirror. This allows us to (i) suppress mirror vibration noise on each axis of the QuAT \cite{Merlet2009, Barrett2016a}, (ii) compensate for gravity-induced Doppler shifts in arbitrary orientations, and (iii) to remove the bias of the classical accelerometers \cite{Cheiney2018}. To evaluate the accuracy of the QuAT, we developed a comprehensive model for systematic effects and we adapted a method for calibrating the triad over a large range of orientations \cite{Yang2012}. In the following, we detail the operation of the QuAT and present results characterizing its short- and long-term performance in different orientations (\ie tilted with respect to gravity) while operating in a static environment.

%------------------------------------------------------------------------------------------------
\subsection*{Vectorial quantum accelerometer}

Acceleration components are measured sequentially by switching between three Mach-Zehnder-type atom interferometers \cite{Kasevich1991} using a $\pi/2-\pi-\pi/2$ sequence of optical Raman pulses to split, reflect, and recombine matter-waves. After an initial cooling and state selection phase, atoms are released into a geodetic free-fall trajectory. During this time, atoms experience a time-varying Doppler shift of $\omega_\mu^{\rm D} = \bm{k}_{\mu} \cdot \bm{g} t$. As each atom interferometer forming the QuAT relies on velocity-sensitive Raman transitions, compensating for this shift is crucial for maintaining the lasers on the two-photon resonance. Atomic gravimeters typically achieve this by chirping the frequency difference between vertical Raman lasers at a quasi-constant rate of $\alpha = k g$, thereby maintaining optimum fringe visibility. In an arbitrary orientation, the projection of gravity on each axis is different and, a priori, not precisely known---necessitating a closed-loop approach to compensate the Doppler shift. Moreover, strong mirror vibrations that blur the interference fringes and further shift the resonance condition \cite{Battelier2016} are a major concern. To address these issues, we implemented a real-time (RT) system (inspired by \Ref \citenum{Lautier2014}) based on a field-programmable gate array (FPGA) that compensates both the frequency and phase of the Raman lasers during the interferometer using input from the classical accelerometers (see Supplementary Materials). This system allows the QuAT to operate over a broad range of orientations and under noisy conditions. To leading order in $T$, the interferometer phase shift along each axis $\mu = x,y,z$ is
\be
  \label{DeltaPhi_mu}
  \Phi_{\mu} = (\bm{k}_{\mu} \cdot \bm{a} - \alpha_{\mu}) T^2 = (k_{\mu} a_{\mu} - \alpha_{\mu}) T^2,
\ee
where $\bm{k}_{\mu} = \frac{1}{c} (\omega_1 + \omega_2) \hat{\bm{\mu}}$ is the effective wavevector of the counter-propagating Raman beams with optical frequencies $\omega_1$ and $\omega_2$, respectively, $\alpha_{\mu}$ is the effective chirp rate at which the Raman frequency is modified between pulses, $T$ is the interrogation time between optical pulses, and $\bm{a}$ is the acceleration vector of the atoms in the body frame (\ie relative to the three orthogonal reference mirrors).

%--------------------------------------------------
\begin{figure}[!t]
  \centering
  \includegraphics[width=0.98\textwidth]{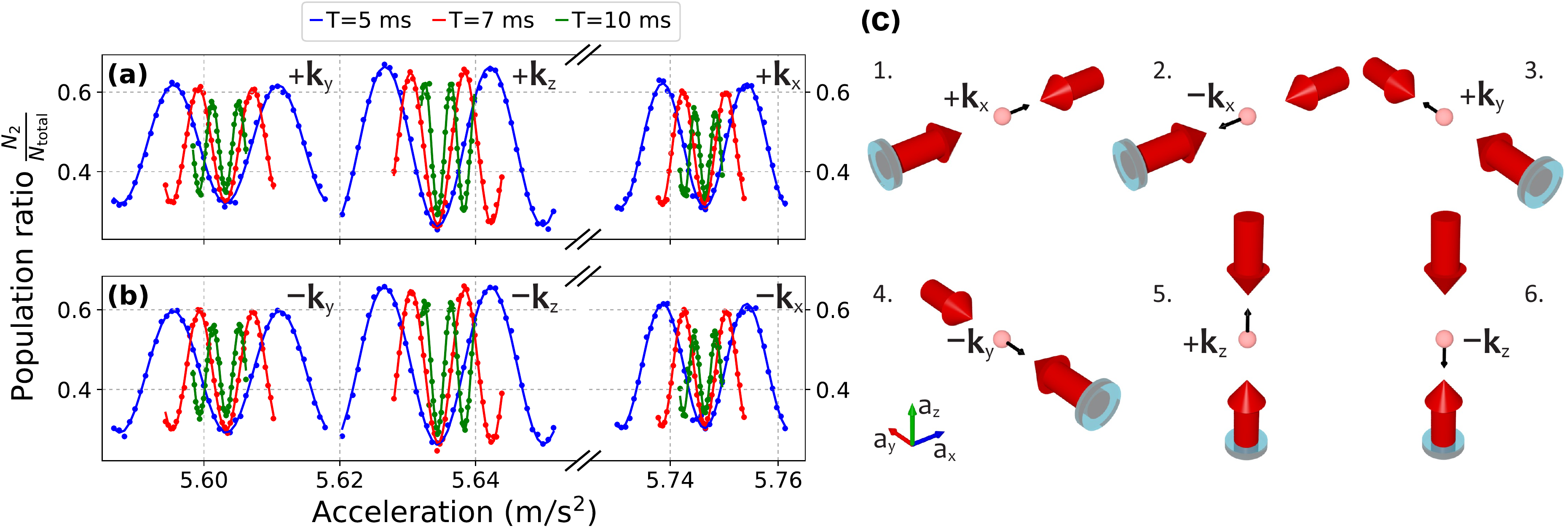}
  \caption{(a,b) Atomic interference fringes obtained along each axis $\mu$ (left: $y$, middle: $z$, and right: $y$) as a function of the laser-induced acceleration $\alpha_{\mu}/k_{\mu}$. Plots (a) and (b) correspond to different momentum transfer directions ($\pm \bm{k}_{\mu}$). Fringes are shown for interrogation times $T = 5$ ms (blue), 7 ms (red), 10 ms (green) at tilt angles of $\theta_x = 54.7^{\circ}$, $\theta_z = 45^{\circ}$. The central fringe common to all $T$ is visible in each set of fringes. (c) Sequence of momentum kicks used during one measurement cycle of the QuAT. Pairs of opposite kicks provide one acceleration component with enhanced rejection of systematic effects.}
  \label{fig:3DFringes}
\end{figure}
%--------------------------------------------------

Figure \ref{fig:3DFringes} illustrates interference fringes obtained along each direction by varying $\alpha_{\mu}$, and the corresponding the measurement sequence. Data are shown for interrogation times $T = 5$, $7$ and $10$ ms, where the fringe spacing in terms of acceleration is given by $2\pi/k_{\mu} T^2$. For each $T$, six fringes are measured by alternating between the axis $\mu$ and the momentum transfer $\pm \hbar \bm{k}_{\mu}$ in an interleaved sequence illustrated in \Fig \ref{fig:3DFringes}(c). Here, a negative (positive) momentum transfer indicates the atoms are kicked toward (away from) the reference mirror. The central fringe for which $\bm{k}_{\mu} \cdot \bm{a} = \alpha_{\mu}$ provides a direct measurement of the acceleration component $a_{\mu}$. The cycle time of our experiment is $T_{\rm cyc} \simeq 1.6$ s (limited by dead time generated by our control system), hence a set of three measurements on orthogonal directions provides the full acceleration vector in under 5 s. We further improve our accuracy by combining measurements with opposite momentum kicks (see Methods), where one full measurement cycle is completed in $6T_{\rm cyc} \simeq 9.6$ s.

The data shown in \Fig \ref{fig:3DFringes} were acquired with the QuAT tilted at approximately $\theta_x = 54.7^{\circ}$, $\theta_z = 45^{\circ}$ such that the gravitational acceleration points along the symmetry axis of the triad. In this orientation, the projection due to gravity along each axis is $a_{\mu} = g/\sqrt{3} \simeq 5.66$ m/s$^2$---providing optimal sensitivity to the full acceleration vector $\bm{a}$. The transit time of the atoms across the Raman beams defines the maximum interrogation time achievable with the present architecture. When the cloud is released from rest at the center of the Raman beams, the maximum interrogation time is limited by the transit time of the atoms across the beams: $\frac{1}{2} a_{\mu} ({\rm TOF} + 2T_{\rm max})^2 = \sqrt{3} w_0$. For a time-of-flight ${\rm TOF} = 20$ ms before the intereferometer, and a beam waist of $w_0 = 11$ mm, we obtain $T_{\rm max} \simeq 31$ ms. To maintain high signal-to-noise ratios on all axes, typically we operate each axis of the QuAT at $T = 10$ ms, where we obtain fringe contrasts of $C_{\mu} = 0.20$, $0.22$, $0.32$, and single-shot acceleration sensitivities of $\delta a_{\mu} = 11.9$, $4.1$, $5.1$ $\mu$g for $\mu = x,y,z$, respectively. The corresponding sensitivity to the vector norm is $\delta |\bm{a}| = \frac{1}{g} \sqrt{\sum_\mu (a_{\mu} \delta a_{\mu})^2} \simeq 7.8$ $\mu$g.

%--------------------------------------------------
\begin{figure}[!t]
  \centering
  \includegraphics[width=0.96\textwidth]{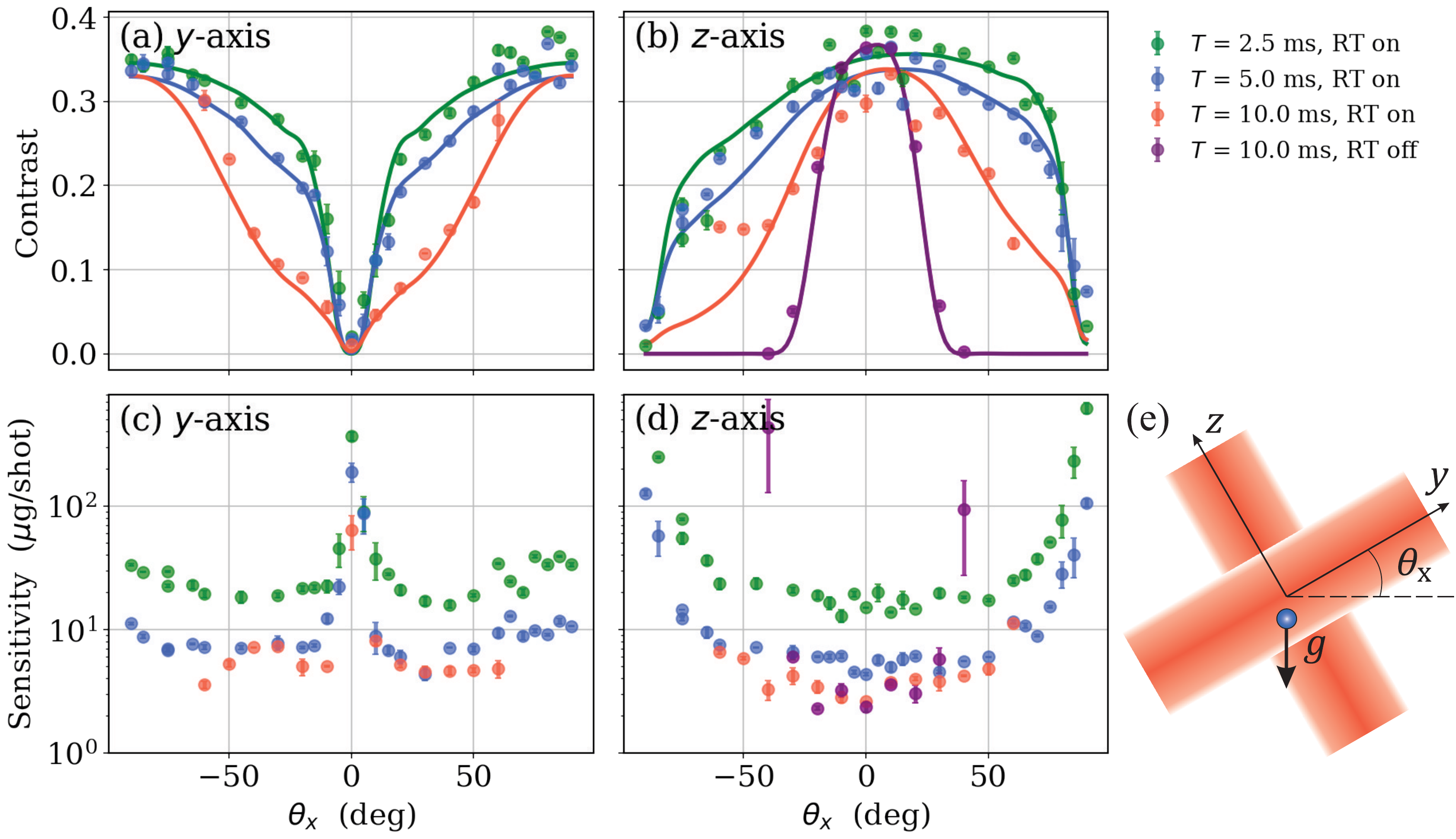}
  \caption{Interference fringe contrast (a,b) and acceleration sensitivity (c,d) along the $y$ and $z$-axes as a function of a random tilt angle $\theta_x$. (e) Geometry of the Raman beams relative to the free-falling atoms. Data were recorded for various interrogation times with the RT system on (green points: $T = 2.5$ ms, blue points: $T = 5$ ms, red points: $T = 10$ ms) and off (purple points: $T = 10$ ms). Error bars correspond to $1\sigma$ uncertainties obtains from sinusoidal fringe fits. Solid lines in (a,b) represent our model of the fringe contrast.}
  \label{fig:Contrast-vs-Tilt}
\end{figure}
%--------------------------------------------------

To illustrate the effect of the RT system, \Fig \ref{fig:Contrast-vs-Tilt} compares the fringe contrast on the $y$- and $z$-axes as a function of the tilt angle $\theta_x$ obtained with this system on and off. Here, the QuAT was oriented such that the $yz$-plane was always vertical, and the tilt about the $x$-axis ($\theta_x$) was varied randomly over $\pm 90$ deg. When the RT system is disabled (purple curve in \Fig \ref{fig:Contrast-vs-Tilt}(b)), the effective chirp rate is fixed such that the Doppler shift is compensated only for the vertical case---resulting in a maximum contrast on $z$ when $\theta_x = 0$. As the $z$-axis is tilted away from the vertical, there is a sharp drop in fringe contrast as the Raman pulses move further off resonance. When enabled, the RT system computes the Doppler shift from the acceleration component normal to the corresponding mirror. It then applies a sequence of phase-continuous frequency steps before each Raman pulse that approximates a chirp with an effective rate $\alpha_{\mu} = k_{\mu} \bar{a}_{\mu}$, where $\bar{a}_{\mu} = \frac{1}{2T} \int_0^{2T} a_{\mu}(t) \dd t$ is the average acceleration during the interferometer. In this case, $\alpha_y \simeq k_y g \sin\theta_x$ and $\alpha_z \simeq k_z g \cos \theta_x$. Figures \ref{fig:Contrast-vs-Tilt}(a) and (b) show that the RT system maintains a near-optimal fringe contrast over $> 50$ deg. Beyond this range, the contrast reduces due to other physical effects as we explain below. We emphasize that despite this drop in contrast, the sensitivity to accelerations remains relatively flat over a broad range of angles. Figures \ref{fig:Contrast-vs-Tilt}(c) and (d) show the acceleration sensitivity per shot: $\delta a_{\mu} = C_{\mu}/(2 \sigma_{\rm N} k_{\mu} T^2)$, where the fringe contrast $C_{\mu}$ and detection noise $\sigma_{\rm N}$ are obtained from fits to interference fringes. At $T = 10$ ms, we obtain $\delta a_{\mu} \lesssim 6$ $\mu$g on each axis even at tilts of 60 deg.

The overall performance of the QuAT is optimal for measuring quasi-static acceleration vectors when it is tilted such that $\bm{a}$ points along its symmetry axis. However, due to limitations of the present architecture, in other orientations the sensitivity of one axis will improve at the expense of the others. This is exacerbated when one axis is nearly parallel with $\bm{g}$. The performance of the QuAT in different orientations is important when considering the case of strapdown navigation \cite{Titterton2004}, where dynamic movements cause both the magnitude and direction of the acceleration vector to change. In an arbitrary orientation, the acceleration sensitivity on any given axis is determined by the longitudinal and transverse motion of the atoms within the Raman beams. The present architecture relies on the Doppler shift of the atoms (generated by longitudinal motion in the beams) to isolate Raman transitions between $\ket{F=1,p_{\mu}}$ and $\ket{F=2,p_{\mu} \pm \hbar k_{\mu}}$. At near-horizontal orientations, the projection of gravity along the beam approaches zero---reducing the Doppler shift such that $\ket{2,p_{\mu}+\hbar k_{\mu}}$ becomes degenerate with several other states (\eg $\ket{2,p_{\mu}-\hbar k_{\mu}}$, $\ket{2,p_{\mu}}$, $\ket{2,p_{\mu}+3\hbar k_{\mu}}$). In this regime, double diffraction processes \cite{Leveque2009, Hartmann2020} and residual velocity-insensitive transitions cause a severe loss in transition probability between our target states. Furthermore, due to the Gaussian intensity profile of the Raman beams, motion in the plane transverse to the Raman wavevector causes a time-varying Rabi frequency during the interferometer. As $T$ increase, the cloud moves toward the edge of the beams where the Rabi frequency approaches zero. These are the two primary effects that produce the contrast variation shown in \Fig \ref{fig:Contrast-vs-Tilt}.

We model the contrast as a function of $\theta_x$ with the RT system on and off, as shown by the solid curves in \Fig \ref{fig:Contrast-vs-Tilt}. Our model (see Methods), which contains only one free parameter---an arbitrary amplitude factor, shows remarkable agreement with the data. When the RT system is disabled, the contrast loss is dominated by the uncompensated Doppler shift $|k_{\mu} g (\cos \theta_x - 1) t|$ that increases dramatically away from vertical ($\theta_x = 0$). When the RT system is enabled, the loss is determined by parasitic diffraction processes, the Gaussian-shaped beam profile, and its finite size. On the $z$-axis, we observe a strong asymmetry in the contrast about $\theta_x$. This is explained by the fact that the initial cloud position is shifted by approximately $-0.8$ mm along the $y$-axis relative to origin where the beams intersect. This results in a longer transit time across the beam for positive tilts ($\theta_x > 0$) compared to negative ones.

%------------------------------------------------------------------------------------------------
\subsection*{Tracking the acceleration vector}

To illustrate the long-term performance of our instrument, we tracked the gravitational acceleration vector over 60 h with the QuAT in a quasi-fixed orientation ($\theta_x = 45^\circ$, $\theta_z = 30^\circ$). Here, we operated the QuAT in closed-loop with the classical accelerometers---creating a hybrid triad (see \Fig \ref{fig:Hybridization} in Methods). Each axis of the QuAT is locked to its central fringe using a $\pm \pi/2$ phase modulation scheme \cite{Merlet2009, Templier2021b} similar to those used in atomic clocks. In this mode, each quantum accelerometer provides a high-accuracy measurement of the corresponding classical accelerometer bias \cite{Cheiney2018}. Data from the three classical accelerometers are simultaneously processed at a rate of 1 kHz, while their biases are sequentially subtracted from each axis at the cycling rate of the experiment ($\sim 0.6$ Hz). In this manner, the hybrid triad retains the best features of both classical and quantum technologies.

\begin{figure}[!t]
  \centering
  \includegraphics[width=0.96\textwidth]{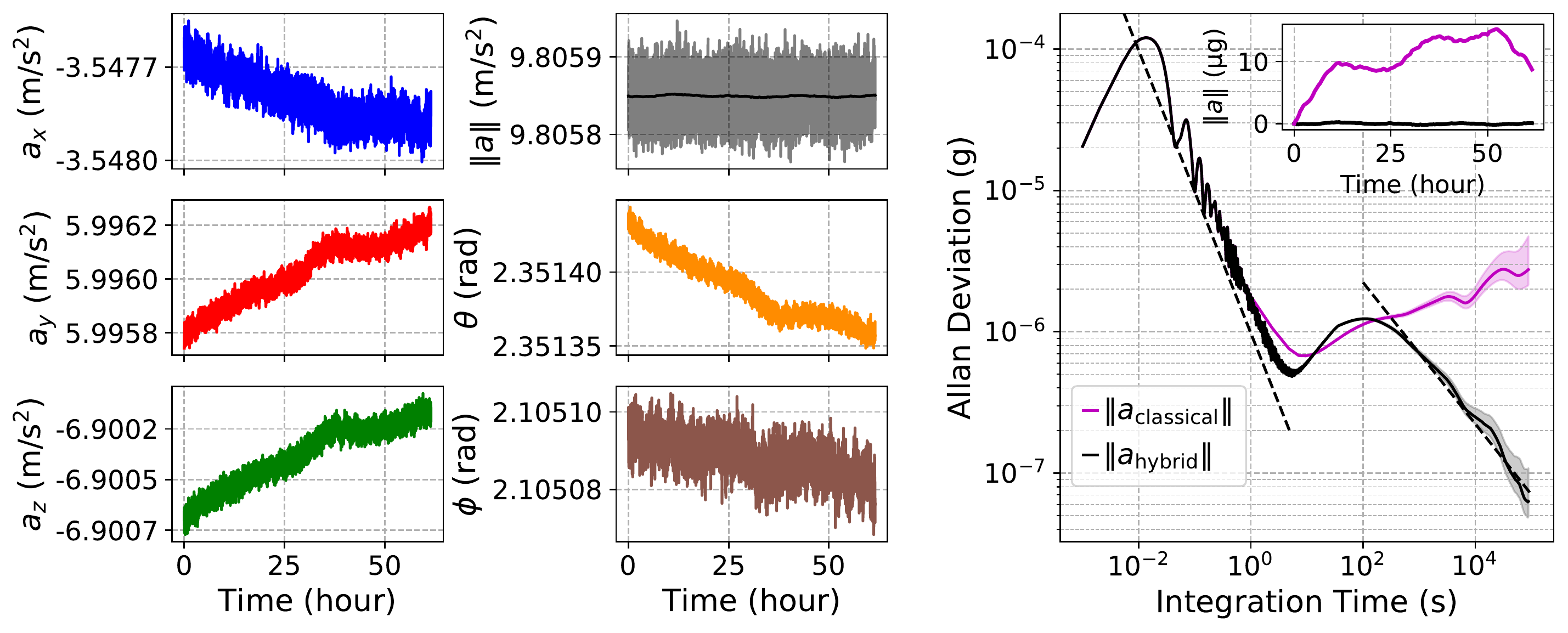}
  \caption{Stability analysis of the hybrid accelerometer triad over 60 h. Here, the interrogation time is $T=10$ ms and the sensor head is tilted at $\theta_x = 45^{\circ}$ and $\theta_z = 30^{\circ}$. (left) Time-series measurements of the hybrid acceleration on each axis in the body frame ($a_x$, $a_y$, $a_z$). These data are displayed at the cycling rate of the experiment. (middle) Same data displayed in spherical polar coordinates with norm $|\bm{a}| = \sqrt{a_x^2 + a_y^2 + a_z^2}$, inclination $\theta = \cos^{-1}(a_z/|\bm{a}|)$, and azimuth $\phi = \mbox{atan2}(a_y,a_x)$. The solid black curve shows a 10-h moving average of $|\bm{a}|$. (right) Allan deviation of acceleration vector norms as a function of integration time $\tau$ for the hybrid (black) and classical (purple) accelerometer triads. Dashed lines represent signals that integrate as $1/\tau$ and $1/\sqrt{\tau}$, respectively. Inset: corresponding time-series produced by the hybrid and classical accelerometer triads.}
  \label{fig:HybridStability}
\end{figure}

Figure \ref{fig:HybridStability} shows an analysis of the hybrid triad output. We represent the acceleration vector in the body frame using both Cartesian ($a_x$, $a_y$, $a_z$) and spherical polar coordinates ($|\bm{a}|$, $\theta$, $\phi$). While both representations are equivalent, the latter best illustrates the physics our instrument is capable of measuring. In Cartesian coordinates, all three vector components exhibit variations at the level of $\sim 5 \times 10^{-4}$ m/s$^2$. These changes could be produced by several different sources (\eg uncompensated bias drifts in the classical accelerometers, changes in relative alignment between quantum accelerometer axes, or drifts of the triad's orientation with respect to $\bm{g}$). In spherical coordinates, it is clear that the vector norm $|\bm{a}|$ remains flat over 60 h within our single-shot measurement noise ($\delta |\bm{a}| \sim 7.6 \times 10^{-5}$ m/s$^2$). As any bias drifts or relative axis misalignments will impact $|\bm{a}|$, we can deduce than these two effects are below the noise level $\delta|\bm{a}|$. This places an upper limit on residual shot-to-shot bias variations of $\delta |\bm{a}|/\sqrt{3} \simeq 4.4 \times 10^{-5}$ m/s$^2$ on each axis, as well as shot-to-shot misalignment variations of $\delta |\bm{a}|/\sqrt{3} g \simeq 4.5$ $\mu$rad. However, the inclination and azimuth angles vary by approximately 70 and 20 $\mu$rad, respectively, over 60 h. This indicates that the triad's orientation is slowly rotating relative to the gravity vector, which produces correlated changes in the Cartesian acceleration components.

%--------------------------------------------------
%--------------------------------------------------

Figure \ref{fig:HybridStability} also shows the Allan deviation of the acceleration vector norm produced by the hybrid and classical accelerometer triads. For timescales of $\tau \le 1$ s, acceleration measurements from both triads are identical and dominated by correlated noise produced by ambient vibrations and the quantization of the analog signals---both of which integrate as $1/\tau$. The classical accelerometer triad reaches a minimum resolution of $7\times 10^{-7}~g$ at 10 s, after which the Allan deviation is limited by the drift of each classical accelerometer bias. In contrast, the output of the hybrid triad features a peak around 100 s (indicative of a feedback loop with an integrator time constant of $\sim 10$ s), followed by a smooth decrease that scales as $1/\sqrt{\tau}$. A fit to this section of the Allan deviation yields a sensitivity of 22 $\mu g/\sqrt{\rm Hz}$, which is limited primarily by the low cycling rate of our instrument. The vector norm of the hybrid accelerometer triad reaches a stability of $6 \times 10^{-8}~g$ after 24 h of integration. In comparison, the classical triad drifts by $\sim 3 \times 10^{-6}~g$ on the same timescale---indicating a 50-fold improvement in vector tracking capability.

%\textcolor{red}{This section has assessed the long-term stability of the hybrid accelerometer. We will now address the static bias of the hybrid sensor with a study of the systematic effects of the atom interferometer and a calibration procedure of the misalignments between axes. }

%------------------------------------------------------------------------------------------------
\subsection*{Systematic effects}
\label{sec:Systematics}

One major advantage of cold-atom-based accelerometers over mechanical ones is their inherent accuracy. Systematic effects in atom interferometers have been studied extensively by several groups \cite{Peters2001, LeGouet2007, Rosi2014, Morel2020, Zhou2021, Barrett2022}, and are well understood. However, the majority of studies to date have considered atoms moving longitudinally along a vertical beam (\eg as in a gravimeter or gradiometer). For the QuAT, the situation is more complex as we must account for effects due to the atom's motion across three orthogonal beams in an arbitrary orientation. We developed a comprehensive model that allows us to vary key system parameters, such as the pulse timing, triad orientation, Raman beam intensities, and the magnetic bias fields. We used this model to characterize the systematic effects for each axis and aid the calibration procedure, which enabled us to evaluate the accuracy of the QuAT.

%--------------------------------------------------
\begin{table}[!th]
    \centering
    \begin{tabular}{lcccccc}
        \hline
        \hline
        Systematic effect      & $a_x^{\rm sys}$ & $a_y^{\rm sys}$ & $a_z^{\rm sys}$ & $|\bm{a}^{\rm sys}|$ & $\delta|\bm{a}^{\rm sys}|$ & Unit \\
        \hline
        Frequency step         &$-57.9(3.8)  $&$ 26.7(4.7)  $&$-26.7(5.2)  $&$ 64.3   $&$ 4.6   $& $\mu g$ \\
        Wavefront curvature    &$  1.69(73)  $&$  2.14(92)  $&$  2.02(87)  $&$ -0.91  $&$ 0.84  $& $\mu g$ \\
        Two-photon light shift &$ -5.53(32)  $&$  6.15(43)  $&$ -4.58(19)  $&$  9.39  $&$ 0.33  $& $\mu g$ \\
        Parasitic lines        &$  0.02(30)  $&$  0.00(30)  $&$ -0.02(18)  $&$  0.00  $&$ 0.26  $& $\mu g$ \\
        Coriolis effect        &$  0.644(60) $&$ -0.077(14) $&$ -0.721(59) $&$  0.000 $&$ 0.049 $& $\mu g$ \\
        One-photon light shift &$  0.000(19) $&$  0.000(12) $&$  0.000(16) $&$  0.000 $&$ 0.016 $& $\mu g$ \\
        Quadratic Zeeman       &$ -0.002(5)  $&$  0.008(3)  $&$ -0.008(3)  $&$  0.010 $&$ 0.004 $& $\mu g$ \\
        RF non-linearity       &$ -0.011     $&$  0.011     $&$ -0.011     $&$  0.020 $&$<0.001 $& $\mu g$ \\
        \hline
        Total                  &$-61.2(3.9)  $&$ 34.9(4.8)  $&$-30.1(5.4)  $&$ 72.8   $&$ 4.7   $& $\mu g$ \\
        \hline
    \end{tabular}
    \caption{Systematic error budget for each axis of the QuAT when tilted along its symmetry axis ($\theta_x = 54.7^{\circ}$, $\theta_z = 45^{\circ}$). Here we assume the axes of the QuAT are mutually orthogonal (\ie no misalignments). Systematic shifts $a_{\mu}^{\rm sys}$ are evaluated assuming path-independent contributions are suppressed by 90\% using $k$-reversal, with $1\sigma$ uncertainties given in parentheses. The systematic shift of the vector norm is $|\bm{a}^{\rm sys}| = \sum_{\mu} a_{\mu} a_{\mu}^{\rm sys}/|\bm{a}|$ (to first order), where $a_x = -a_y = a_z \simeq -5.66$ m/s$^2$ in this orientation. Other parameters: TOF $= 20$ ms, $2\tau_{\mu} = (12,11,10)$ $\mu$s, $T = 10$ ms; Rabi frequencies for velocity-sensitive transitions $\Omega_{x,1} = 0.299(75)$ rad/$\mu$s, $\Omega_{x,3} = 0.138(20)$ rad/$\mu$s, $\Omega_{y,1} = 0.273(55)$ rad/$\mu$s, $\Omega_{y,3} = 0.165(51)$ rad/$\mu$s, $\Omega_{z,1} = 0.316(75)$ rad/$\mu$s, $\Omega_{z,3} = 0.200(78)$ rad/$\mu$s; Rabi frequencies for velocity-insensitive transitions $\Omega_{\mu,j}^{\rm co} = 0.68(10) \Omega_{\mu,j}$, $B$-field strengths $B_x = 139.41(38)$ mG, $B_y = 137.70(27)$ mG, $B_z = 147.71(73)$ mG; $B$-field gradients $\partial_x B_x = -0.045(19)$ G/m, $\partial_y B_y = 0.211(11)$ G/m, $\partial_z B_z = -0.204(11)$ G/m; cloud temperature $\mathbb{T} = 3.5(1.0)$ $\mu$K; initial cloud position and velocity uncertainties: $\delta \mu = 0.2$ mm, $\delta v_{\mu} = 3$ mm/s; curvature of Raman wavefronts $R_x = 5.0(1.7)$ km, $R_y = 4.2(1.4)$ km, $R_z = 4.2(1.4)$ km.}
    \label{tab:Systematics}
\end{table}
%--------------------------------------------------

Table \ref{tab:Systematics} provides a budget of systematic errors when the triad is tilted at 54.7$^{\circ}$ along its axis of symmetry, where all acceleration components have the same projection due to gravity. The largest systematic effect along any given axis is induced by our frequency-step protocol for the RT compensation of the Doppler effect. We apply a series of phase-continuous frequency steps to the Raman frequency that mimics a true frequency chirp between pulses, but maintains a constant frequency during the pulses. This creates a slight imbalance between the kinematic phase due to atomic motion and the phase imprinted by the Raman laser---resulting in a phase shift proportional to the difference between the Rabi frequencies at the beamsplitter pulses:
\be
  \phi_{\mu}^{\rm sys} \simeq -k_{\mu} a_{\mu} T (\pi/2 - 1) (\Omega_{\mu,3} - \Omega_{\mu,1}) (2\tau_{\mu}/\pi)^2.
\ee
Here, $\Omega_{\mu,j}$ is the effective Rabi frequency along axis $\mu$ during Raman pulse $j$. In previous work \cite{Templier2021b}, we evaluated this systematic and its coupling to parasitic laser lines on the $z$-axis. When oriented vertically, the shift due to the frequency-step protocol is $< 1$ $\mu g$ at $T = 10$ ms. However, this effect is exacerbated in tilted configurations because the atoms experience a larger variation in the Rabi frequency as they transit the Raman beams. This also increases other systematic effects, such as the wavefront curvature \cite{Karcher2018}, two-photon light shift \cite{Gauguet2008}, and the Coriolis effect (although the latter cancels in the vector norm\footnote{The first-order shift on the vector norm due to Coriolis acceleration is proportional to $\sum a_\mu a_{\mu}^{\rm Coriolis}$, which cancels if the initial launch velocity of the atom cloud is null. The uncertainty in this shift listed in Table \ref{tab:Systematics} is due to the initial velocity uncertainties.}). Both the frequency-step shift and the two-photon light shift are proportional to the intensity of the Raman beams, hence they can be suppressed in future iterations by subtracting measurements taken at two different intensities \cite{Louchet-Chauvet2011a}. This would reduce both effects by more than a factor of 10 at the expense of increased measurement time. Replacing the frequency-step protocol with a phase-continuous chirp would eliminate this systematic entirely. When coupled with minor improvements to the wavefront curvature, these enhancements would enable us to reduce the systematic uncertainty on each axis below 100 n$g$.

%--------------------------------------------------
\begin{figure}[!b]
  \centering
  \includegraphics[width=0.96\textwidth]{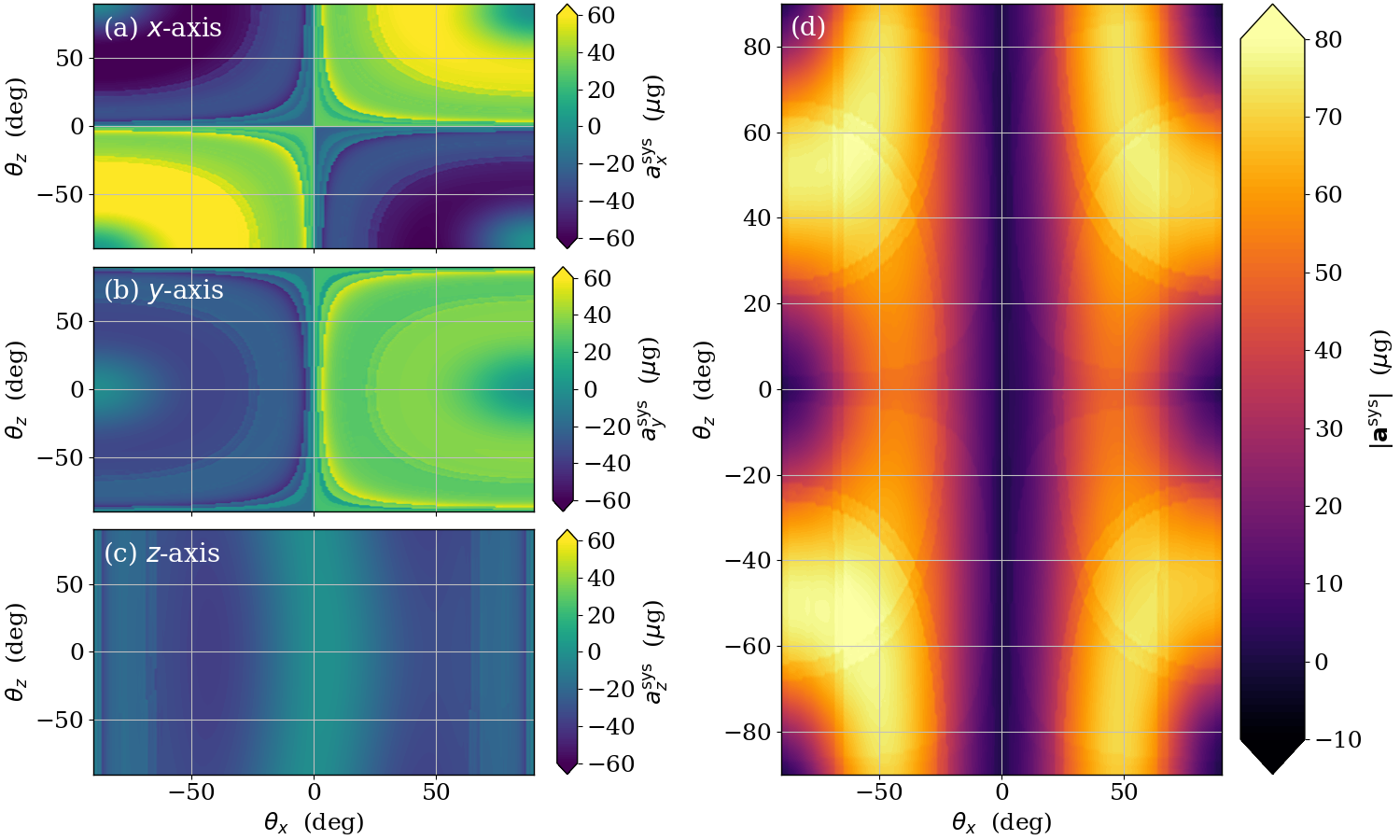}
  \caption{Model of systematic shifts as a function of tilt angles $\theta_x$ and $\theta_z$. The total shift $a_{\mu}^{\rm sys}$ is shown for each axis of the QuAT (a--c), as well as the vector norm $|\bm{a}^{\rm sys}|$ (d). Experimental parameters are the same as for Table \ref{tab:Systematics}. All graphs show the mean of systematic shifts with opposite momentum transfer: $\frac{1}{2} [a_{\mu}^{\rm sys}(+\bm{k}_{\mu}) + a_{\mu}^{\rm sys}(-\bm{k}_{\mu})]$.}
  \label{fig:SystematicsModel}
\end{figure}
%--------------------------------------------------

Figure \ref{fig:SystematicsModel} illustrates the complex dependence of the systematics on the orientation of the QuAT. As discussed above, the topology is governed primarily by three effects: our frequency-step protocol, the curvature of Raman wavefronts, and the two-photon light shift. All of these effects increase dramatically when the beams are near horizontal. The frequency-step and wavefront systematics are dominated by the atoms sampling different regions of the Raman beam profile, while the two-photon light shift is strongly affected by the separation between neighbouring Raman transitions. This effect produces regions in \Fig \ref{fig:SystematicsModel} with sharp changes in contrast, where velocity-insensitive or magnetically-sensitive transitions are near-resonant with the primary $\ket{1,p_{\mu}} \to \ket{2,p_{\mu} \pm \hbar k_\mu}$ transition. To evaluate the effect on the acceleration vector norm, we sum the systematic shifts on each axis weighted by the corresponding projection on norm. The results shown in \Fig \ref{fig:SystematicsModel}(d) indicate that the systematic shift on the norm is largest in regions where the projections from each axis are similar in magnitude (\eg near $\theta_x = \pm 54.7^{\circ}$, $\theta_z = \pm 45$), which is consistent with our expectations.

%------------------------------------------------------------------------------------------------
\subsection*{Calibration of the quantum accelerometer triad}
\label{sec:Calibration}

\begin{figure}[!hb]
  \centering
  \includegraphics[width=0.96\textwidth]{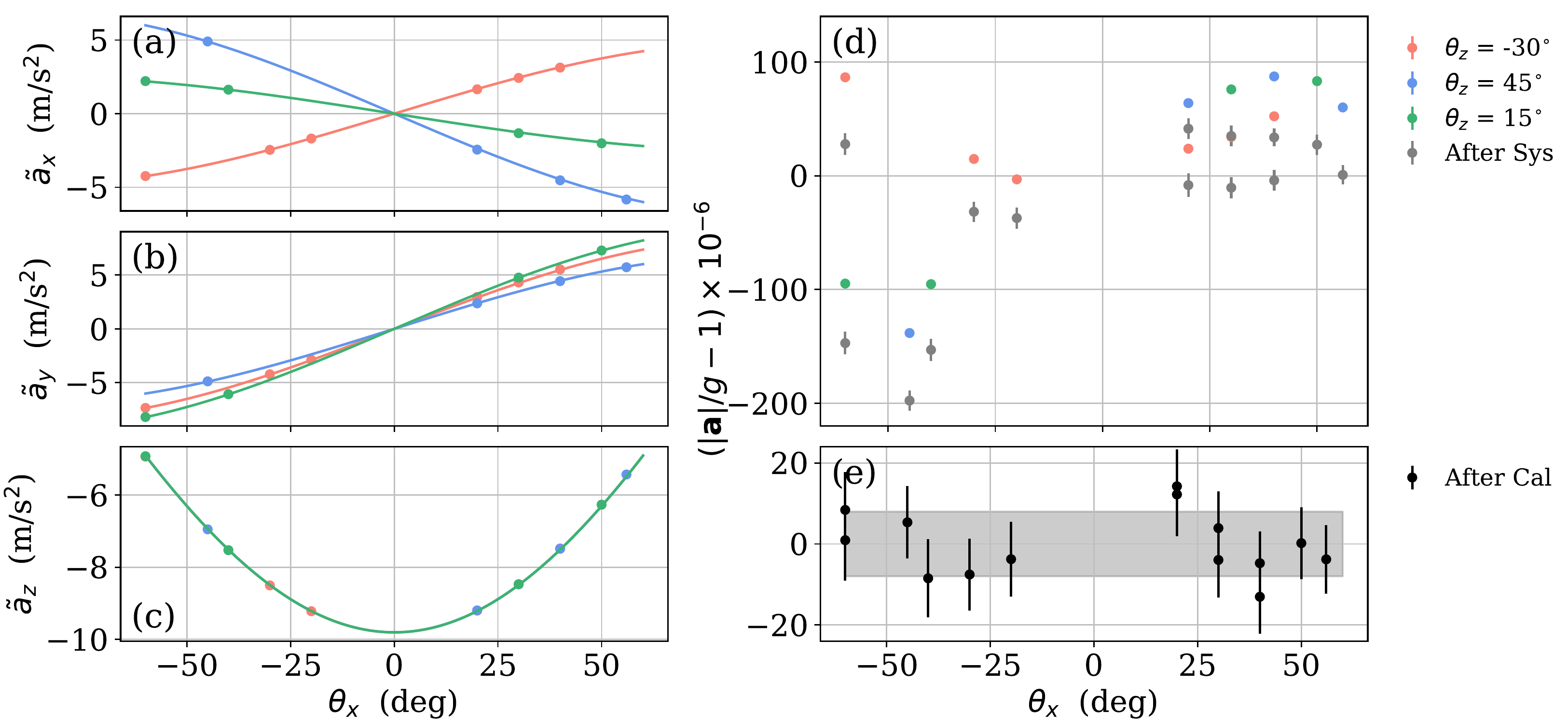}
  \caption{Data used to calibration the QuAT. The acceleration vector is sampled at 14 independent orientations (labeled by tilt angles $\theta_x$ and $\theta_z$) using an interrogation time of $T = 10$ ms. (a-c) Measured acceleration components $\tilde{a}_x$, $\tilde{a}_y$, and $\tilde{a}_z$ for each orientation. Solid lines correspond to expected variations with $\theta_x$. (d) Error in the vector norm before (colored) and after (gray) the subtraction of systematics. Error bars for the raw data are smaller than the points. (e) Zoom of the vector norm error after the calibration procedure. The gray band indicates the remaining rms spread of $\pm 7.7$ $\mu g$.}
  \label{fig:QuAT_Calibration}
\end{figure}

Misalignments between axes and variations in the accelerometer scale factors directly affect the accuracy of both the magnitude and direction of the acceleration vector. The triad axes are defined by the effective wavevectors $\bm{k}_{\mu}$, which are normal to the surface of the corresponding retro-reflection mirror. As these mirrors are oriented at approximately $90^{\circ}$ relative to each other, the error in the acceleration vector scales to first order with misalignment angle. These misalignments are challenging to measure and stabilize at levels below 10 $\mu$rad due to mechanical strain and thermal expansion \cite{Xu2017}. In addition to misalignments, the accuracy of the QuAT is affected by the scale factor of each quantum accelerometer ($S_{\mu} \simeq k_{\mu} T^2$). These quantities depend on the absolute frequency of our Raman lasers, the angle between the incident and retro-reflected Raman beams, the timing between laser pulses, and the beam intensity sampled by the atoms. To account for imperfections in the alignment and scale factors of the QuAT, we model its output in an arbitrary static orientation as follows:
\be
  \label{TriadModel}
  \begin{bmatrix} \tilde{a}_x \\ \tilde{a}_y \\ \tilde{a}_z \end{bmatrix} =
  \begin{bmatrix}
    \kappa_x & 0   & 0 \\
    0   & \kappa_y & 0 \\
    0   & 0   & \kappa_z
  \end{bmatrix}
  \begin{bmatrix}
    1 & 0 & 0 \\
    \lambda_{yx} & 1 & 0 \\
    \lambda_{zx} & \lambda_{zy} & 1
  \end{bmatrix}
  \begin{bmatrix} a_x \\ a_y \\ a_z \end{bmatrix} +
  \begin{bmatrix} a_x^{\rm sys} \\ a_y^{\rm sys} \\ a_z^{\rm sys} \end{bmatrix},
\ee
where $\tilde{a}_{\mu}$ is a measured acceleration component, $\kappa_{\mu}$ is a relative scale factor, $\lambda_{\mu\nu}$ is the misalignment factor between axes $\mu$ and $\nu$, and $a_{\mu}^{\rm sys}$ is a systematic bias. Here, $a_{\mu}$ is the gravitational acceleration projected onto a mutually-orthogonal Cartesian frame, hence they obey $a_x^2 + a_y^2 + a_z^2 = g^2$ where $g = 9.805642$ m/s$^2$ is the local gravitational acceleration. To estimate these parameters, we measured the gravity vector at several different orientations, as shown in \Fig \ref{fig:QuAT_Calibration}. We then subtract orientation-dependent systematic shifts, and fit our triad model [\Eq \eqref{TriadModel}] to the resulting data using an iterative minimization algorithm \cite{Yang2012}. Table \ref{tab:QuAT_Calibration} summarizes the parameters that best fit the data---indicating misalignment angles up to $\lambda_{\mu\nu} \simeq 200$ $\mu$rad between axes\footnote{The misalignment between the classical accelerometers and the quantum ones is absorbed into their respective scale factors. Effectively, the hybrid system forces these two triads to be aligned at the expense of slightly smaller classical accelerometer scale factors.} and relative scale factors $\kappa_{\mu}$ within 50 parts per million of unity. These differ from unity primarily due to the finite pulse lengths $\tau_{\mu}$ and the non-ideal Rabi frequencies $\Omega_{\mu,j}$, which affect the atom interferometer scale factors according to \cite{Bonnin2015, Templier2021b}
\be
  S_{\mu} = k_{\mu} (T + 2\tau_{\mu}) \left[T + \frac{1}{\Omega_{\mu,1}} \tan\left( \frac{\Omega_{\mu,1} \tau_{\mu}}{2} \right) + \frac{1}{\Omega_{\mu,3}} \tan\left( \frac{\Omega_{\mu,3} \tau_{\mu}}{2} \right) \right].
\ee
To convert our atom interferometer phase measurements to accelerations, we used a scale factor with ideal Rabi frequencies: $S_{\mu}^{\rm ideal} = k_{\mu} (T + 2\tau_{\mu}) (T + 4\tau_{\mu}/\pi)$. For our experimental parameters, we find that the ratio between these scale factors varies between $S_{\mu}/S_{\mu}^{\rm ideal} \simeq 0.999940 - 0.999980$, which is consistent with the results shown in Table \ref{tab:Systematics}.

%--------------------------------------------------
%--------------------------------------------------
%--------------------------------------------------
\begin{table}[!bt]
    \centering
    \begin{tabular}{cc}
        \hline
        \hline
        Scale factors & Misalignments \\
        \hline
        $\kappa_x = 0.9999563(41)$ & $\lambda_{yx} = +1.926(42) \times 10^{-4}$ rad \\
        $\kappa_y = 0.9999702(25)$ & $\lambda_{zx} = +1.967(43) \times 10^{-4}$ rad \\
        $\kappa_z = 0.9999619(18)$ & $\lambda_{zy} = -1.359(25) \times 10^{-4}$ rad \\
        \hline
    \end{tabular}
    \caption{QuAT model parameters resulting from the calibration procedure. The $1\sigma$ uncertainties are provided in parentheses. The misalignment angles are determined to approximately 4 $\mu$rad---providing a direct metric of the vector pointing accuracy.}
    \label{tab:QuAT_Calibration}
\end{table}
%--------------------------------------------------
Figure \ref{fig:QuAT_Calibration} illustrates the effect of the triad calibration. Vector accelerations were recorded over tilt angles ranging from $\theta_x = -55^{\circ}$ to $+55^{\circ}$ and $\theta_z = -30^{\circ}$ to $+45^{\circ}$ (see \Fig \ref{fig:3DAccelDesign} for reference). This large range of orientations causes the acceleration components to vary over roughly $\pm 5$ m/s$^{2}$. Each orientation of the QuAT was set by carrying out a sequence of two extrinsic rotations: one about the vertical $z$-axis with an angle $\theta_z$, followed by a rotation about the horizontal $x$-axis by $\theta_x$. This sequence is described by the following transformation of the gravity vector
\be
  \begin{bmatrix} a_x \\ a_y \\ a_z \end{bmatrix} =
  \begin{bmatrix}
    \cos\theta_z & -\sin\theta_z\cos\theta_x & +\sin\theta_z\sin\theta_x \\
    \sin\theta_z & +\cos\theta_z\cos\theta_x & -\cos\theta_z\sin\theta_x \\
    0            &  \sin\theta_x             &  \cos\theta_x
  \end{bmatrix}
  \begin{bmatrix} 0 \\ 0 \\ -g \end{bmatrix},
\ee
The measured accelerations reflect this dependence on $\theta_x$ and $\theta_z$, as shown in \Figs \ref{fig:QuAT_Calibration}(a--c). However, small imperfections in the triad lead to significant errors in the vector norm. Figure \ref{fig:QuAT_Calibration}(d) shows that, in some orientations, the uncalibrated QuAT produces errors as large as 140 $\mu g$. However, these errors are effectively corrected using the calibration procedure. We test the calibration by inverting \Eq \eqref{TriadModel} with the parameters in Table \ref{tab:QuAT_Calibration} to determine the true acceleration vectors. Prior to calibration, the root-mean-squared (rms) error in the vector norm is 74 $\mu g$. After the calibration procedure, the rms error decreases by almost an order of magnitude to 7.7 $\mu g$. We anticipate further improvements can be gained by compensating the measurements for the variation of interferometer scale factors with orientation.

%================================================================================================
\section*{Discussion}
\label{sec:Discussion}
% The discussion describes the conclusions that can be drawn from the results, as well as the significance and implications of the research. A paragraph discussing the limitations of the study should be included and any issues that will need to be addressed before application to animal, human, or environmental health should also be described.

We have achieved full 3D tracking of the acceleration vector with a compact hybrid quantum accelerometer triad (QuAT) and demonstrated a 50-fold improvement in the long-term bias stability over navigation-grade accelerometers. Our instrument combines sequential interrogation of three single-beam atom interferometers coupled with a classical accelerometer triad. Fusion of all data from the QuAT provides the quantum advantage in short-term accuracy and low long-term bias drift together with the high data rate required to track the acceleration.

Accurate positioning and navigation often requires the fusion of data from global navigation satellite systems (GNSS) and autonomous inertial navigation systems (INS). The latter heavily relies on triads of accelerometers and gyroscopes, where the attitude and position of a moving body is determined by integrating the equations of motion \cite{Titterton2004}. The accuracy of an INS is limited by the bias stability of the inertial sensors \cite{Groves2013, Lefevre2014}, as well as the knowledge of the local gravitational field. Taking advantage of the quantum nature of our sensor, its high sensitivity and low bias stability can resolve all these challenges \cite{Jekeli2005, Cheiney2018}. It is sensitive to acceleration resulting from motion (AC) as well as from gravity (DC), and exhibits a long-term stability of 60 n$g$ (60 $\mu$Gal). The short-term (AC) sensitivity of the QuAT is mostly limited by the classical accelerometers to 100 $\mu g$ at an interrogation time of $T = 10$ ms. Our model of systematic effects combined with our calibration procedure leads to a DC accuracy of 7.7 $\mu$g (7.7 mGal) on the vector norm and a pointing accuracy of $\sim 4$ $\mu$rad relative to each axis.

The results presented here demonstrate the full potential of matter-wave inertial sensors for future quantum-aided navigation, either by using the QuAT output to directly determine a vehicle's position, or by providing strapdown operation for gravity mapping \cite{Bidel2018, Bidel2020} or gravity matching-aided navigation \cite{Gao2021, Wang2022}. They can also be used to reduce the bias drift on the local acceleration readings and thus relieve the constraint on Schuler, Foucault, or other Earth-periodic oscillation errors \cite{Zhao2015}. %\textcolor{red}{Our work is an essential step towards these onboard applications and further developments is still to be made for an implementation in the field, including a real time rotation compensation to maintain the contrast and reduce rotation induced phase noise and an improvement of the correlation between the classical and the quantum sensors in presence of significant accelerations. Moreover, navigation will require the hybridization with gyroscopes to benefit from the improved acceleration signal. }

The pointing accuracy of the QuAT, together with its long-term stability provides a promising alternative for high-resolution tidal tilt measurements \cite{Liu2022}. An correlated array of such accelerometer triads, or their combination with precision rotation measurements, can render angle measurements immune to external noise and improve our understanding of ground motion---representing a major stake for seismology. Long-term angle monitoring can provide knowledge of translational and rotational motion that could significantly improve seismic inverse models for the Earth’s structure \cite{Schmelzbach2018}, allow for full seismic signal reconstruction and modeling \cite{VanRenterghem2017}, or help characterize earthquake sources \cite{Reinwald2016} and their points of origin \cite{Li2017}.

%================================================================================================
\section*{Materials and Methods}
% The materials and methods section should provide sufficient information to allow replication of the results. Begin with a section titled Experimental Design describing the objectives and design of the study as well as pre-specified components.
% In addition, include a section titled Statistical Analysis at the end that fully describes the statistical methods with enough detail to enable a knowledgeable reader with access to the original data to verify the results. The values for N, P, and the specific statistical test performed for each experiment should be included in the appropriate figure legend or main text.

Here we describe the apparatus, including the multi-axis sensor head, laser source, and a RT vibration control system. We also discuss our models for the fringe contrast and systematic effects.

%------------------------------------------------------------------------------------------------
\subsection*{Sensor head}
\label{sec:Sensor}

To reduce the size and complexity of our apparatus, we adopted a three-beam architecture where retro-reflected light along each axis can be used for trapping, cooling, manipulating, and detecting the atoms. This requires independent control of the optical power and polarization on each axis. Our design includes three optical collimators that expand the light to a $1/e^2$ diameter of $\sim 22$ mm, and three retro-reflection mirrors to which we attach mechanical accelerometers for hybridization purposes. We use navigation-grade pendulous rebalance accelerometers manufactured by Thales (J192AAM on the $x,y$-axes, EMA 1000-B1 on the $z$-axis) which feature a high-sensitivity and high-bandwidth response. They feature an intrinsic bias between 50-180 $\mu g$, a scale factor variation of 120 ppm/$^{\circ}$C, a flat response from DC to approximately 300 Hz, and a monotonically decreasing sensitivity up to $\sim 1$ kHz. For the hybridization process, we acquire raw data from these sensors at 5 kHz. The output of the hybrid sensor is digitally filtered prior to being streamed to disc at a data rate of 1 kHz. These accelerometers also feature a magnetic shield, which is crucial for reducing their sensitivity to the relatively strong, pulsed magnetic fields produced by nearby MOT coils.

All of the optics are fixed directly to a forged titanium vacuum chamber---forming the rigid triad shown in \Fig \ref{fig:3DAccelDesign}(b). Magnetic bias and gradient coils required for atom trapping and interferometry are wound within circular grooves machined directly on the $x$- and $y$-axes of the chamber. Two pairs of square coils fixed outside the chamber provide both a bias field and a gradient field on the $z$-axis. The entire vacuum system is mounted within a single-layer $\mu$-metal shield to stabilize the $B$-field experienced by the atoms when the system is rotated. Excluding the magnetic shield, the volume of the sensor head is approximately 45 L and weighs 40 kg. This includes the vacuum chamber, ion pump, fiber collimators, and detectors.

An all-fibered optical bench at 780 nm, mounted within the shield, includes polarizing cube-based fiber splitters (Thorlabs PBC780PM-APC), and a $1 \times 4$ micro-optic fiber switch (Leoni EOL 1x4) to alternate between cooling on all axes and interferometry on each axis. The light is produced by a unique dual-frequency laser source developed by iXblue (Modbox laser, 6U, 19'' rack mounted) based on telecom components at 1560 nm and wavelength conversion to 780 nm in a periodically-poled lithium niobate waveguide (NTT WH-0780-000-F-B-C). This laser source, which utilizes a unique optical IQ modulator to derive all required frequencies, is described in detail elsewhere \cite{Templier2021a, Templier2021b}. An oven-controlled liquid-crystal retarder (LCR) (Thorlabs LCC1111T-B) is placed at the output of each collimator which allows us to switch between the polarizations required for cooling (circular) and coherent Raman transitions (linear) during the measurement sequence. We maintain the LCRs at a temperature of $\sim 53^{\circ}$ to stabilize the switching time between polarization states ($\sim 500$ $\mu$s from circular to linear). A $\lambda/4$-wave plate is also mounted in front of each retro-reflection mirror to flip the polarization on the return path [see \Fig \ref{fig:3DAccelDesign}(a)]. This ensures that counter-propagating beams for cooling and trapping have opposite circular polarization, and the Raman beams have perpendicular linear polarization---allowing us to minimize parasitic velocity-insensitive transitions.

From the vapor-loaded 3D magneto-optical trap (MOT) we obtain approximately $10^8$ $^{87}$Rb atoms in 250 ms. This sample is subsequently cooled to approximately 3 $\mu$K in a gray molasses on the D2-transition at 780 nm \cite{Rosi2018}. We prepare these atoms in the magnetically-insensitive $\ket{F=1, m_F=0}$ state by: (i) initially pumping them to $\ket{F=1}$; (ii) applying a quantizing $B$-field of 140 mG along a given axis ($\mu = x, y, z$); and (iii) removing atoms in the $\ket{F=1, m_F=\pm 1}$ states through a coherent optical transfer to $\ket{F=2}$, followed by a near-resonant push pulse \cite{Templier2021a}. To measure different acceleration components, we switch between axes sequentially using the independent Raman beams and corresponding pairs of Helmholtz coils. At this point, the two Raman beams are detuned by $\Delta_{\rm R} = -880$ MHz from the $\ket{F' = 2}$ excited state in $^{87}$Rb and we apply a $\pi/2-\pi-\pi/2$ sequence of Raman pulses separated by interrogation time $T$. This is followed by a fluorescence detection phase where a sequence of near-resonant pulses is applied along the $z$-axis to measure the ratio of atoms in $\ket{F=2}$. Our detection system is composed of two photodiodes with large fields of view to measure the fluorescence from the atoms over a broad range of orientations. The effective detection volume of this system is a sphere with a diameter of approximately 26 mm. This enables us to observe fluorescence up to flight times of $\sim 52$ ms. The detection signal is digitized and processed to determine the acceleration component on a given axis.

This sequence of operations occurs once per measurement cycle ($T_{\rm cyc} \simeq 1.6$ s), which is limited by software dead time \cite{Keshet2013}. A minimum of two measurement cycles are required to obtain one acceleration component (one on each side of the interference fringe at $\pm \pi/2$), hence the full acceleration vector can be obtained in 6 measurement cycles ($\sim 9.6$ s). However, in practice we use 6 cycles with a momentum transfer direction $+\bm{k}_{\mu}$ interleaved with 6 cycles using $-\bm{k}_{\mu}$ in order to reject path-independent systematic biases on each axis. Hence, we construct the full acceleration vector once every 12 cycles ($\sim 19.2$ s).

%------------------------------------------------------------------------------------------------
\subsection*{Rotation platform}
\label{sec:Platform}

The QuAT is installed on a manual three-axis rotation platform that can be rotated continuously by $360^{\circ}$ about three independent axes. This platform was initially designed for small loads and had to be modified to accommodate the QuAT. To compensate for the mass of the sensor head (40 kg) and the magnetic shield (40 kg), we constructed an 80 kg ballast system below the primary platform. This keeps the center of mass near the center of rotation---minimizing the torque on the bearings. However, due to mechanical constraints of this system, the horizontal $x$-axis was limited to rotation angles of $\theta_x \in [-90^{\circ}, +90^{\circ}]$, and the vertical $z$-axis to angles $\theta_z \in [-180^{\circ}, +180^{\circ}]$ in steps of $15^{\circ}$. This freedom allows us to project different amounts of gravity along each axis, which creates unique Doppler shifts due to the atom’s free fall and breaks the degeneracy between $\pm \hbar \bm{k}_{\mu}$ momentum transfers. The general form of the Doppler shift is
\be
  \label{omegaD}
  \omega_{\mu}^{\rm D}(t) = \pm k_{\mu} \int_0^t a_{\mu}(t') \dd t' \approx \pm k_{\mu} \bar{a}_{\mu} t,
\ee
where the explicit time-dependence of $a_{\mu}(t)$ accounts for the motion of the atoms relative to the surface of the mirror during the interferometer sequence, and the $\pm$ corresponds to opposite momentum transfers. The approximate form in \Eq \eqref{omegaD} is valid when the variations in acceleration during the interferometer are negligible. Here, $\bar{a}_{\mu} = \frac{1}{2T} \int_0^{2T} a_{\mu}(t') \dd t'$ is the average acceleration during the interferometer. At any time $t$ the resonance condition is determined by the two-photon detuning $\delta_{\mu}$ given by
\be
  \label{delta}
  \delta_{\mu}(t) = \Delta\omega_{\mu}^{\rm R}(t) - \omega_{\rm HF} - \omega_{\mu}^{\rm D}(t) - \omega_{\rm rec},
\ee
where $\Delta\omega_{\mu}^{\rm R}(t)$ is the difference between optical Raman frequencies, $\omega_{\rm HF} \simeq 2\pi \times 6.834$ GHz is the ground state hyperfine splitting, and $\omega_{\rm rec} \simeq 2\pi \times 15.1$ kHz is the recoil frequency for $^{87}$Rb. For clarity, we have ignored smaller frequency shifts (\eg due to the AC stark effect) in \Eq \eqref{delta}. The resonance condition ($\delta_{\mu} = 0$) must be satisfied to optimize the transfer efficiency between the two target states ($\ket{1,p}$ and $\ket{2,p \pm \hbar k_{\mu}}$). Hence, to maximize the fringe contrast of the atom interferometer at interrogation times $T > \Omega_{\mu}/|k_{\mu} \bar{a}_{\mu}|$, one must compensate the time-varying Doppler shift experienced by the free-falling atoms. This is typically achieved by applying a phase-continuous chirp to the Raman frequency: $\Delta\omega_{\mu}^{\rm R}(t) \simeq \omega_{\rm HF} + \omega_{\rm rec} + \alpha_{\mu} t$, where $\alpha_{\mu} = \pm k_{\mu} \bar{a}_{\mu}$. This maintains the resonance condition as the atoms accelerate toward or away from the mirror. In a fixed orientation, the chirp rate is constant for each axis and can be determined experimentally by locating the central fringe for which the total interferometer phase is zero [see \Eq \eqref{DeltaPhi_mu}]. However, if the orientation of the triad is not precisely known, determining $\alpha_{\mu}$ with this method can be very time consuming. Furthermore, if the system is rotating or undergoing translational acceleration (\ie if the triad is mobile), $\alpha_{\mu}$ needs to be updated on a shot-to-shot basis. In the most extreme case, when mirror accelerations exceed $a_{\mu}(t) > \Omega_{\mu}/k_{\mu} T$, the Doppler shift cannot be approximated as linear \cite{Barrett2016a, Battelier2016} and must be compensated \emph{during} the interferometer sequence. All of these scenarios can be addressed with a real-time (RT) solution. We describe in detail our RT system in the Supplementary Materials.

%------------------------------------------------------------------------------------------------
\subsection*{Hybridization of the quantum and classical accelerometers}
\label{sec:Hybridization}

We hybridize the quantum and classical triads by establishing a feedback loop between the quantum phase measurements and the classical accelerometer signals. The hybridization can be understood as follows: the surfaces of the three retro-reflection mirrors define the 3D reference frame for the atoms, and the relative motion of the triad compared to the free-falling atoms is recorded by the classical accelerometers. Because of ambient vibrations, the reference frame shakes and by itself produces phase noise on our interference fringes. To solve this issue, our RT system effectively stabilizes the Raman beams to the atoms' free-falling frame by correcting their relative frequency and phase. This allows us to suppress vibration noise on each axis of the QuAT, as well as to periodically measure the bias of each classical accelerometer. These biases are then subtracted from the continuous output of the classical accelerometers once per measurement cycle and fed back to the QuAT one axis at a time---closing the feedback loop. In this way, we generate an ultra-stable, high-bandwidth hybrid accelerometer triad.

%--------------------------------------------------
\begin{figure}[!t]
  \centering
  \includegraphics[width=0.96\textwidth]{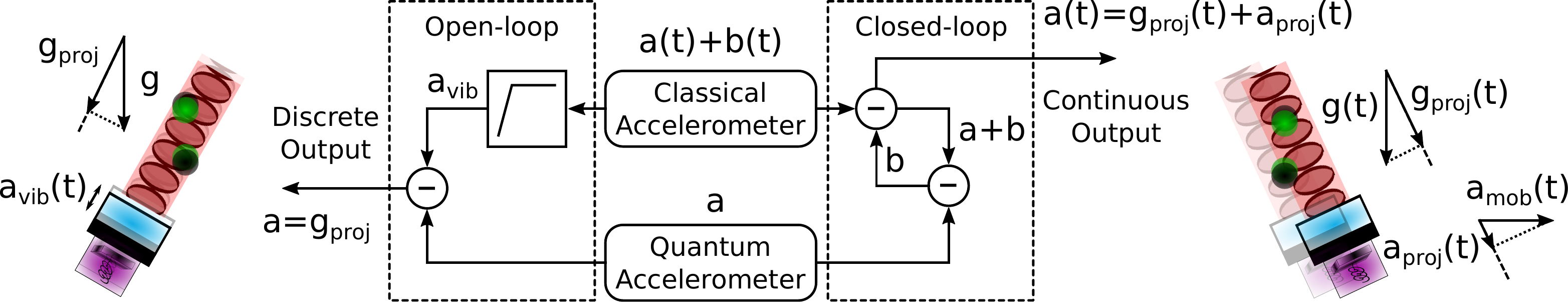}
  \caption{Hybridization schemes between the quantum and classical accelerometers. The open-loop scheme on the left depicts how the filtered classical accelerometer is used to correct vibrations of the reference mirror $a_{\rm vib}$ for the quantum accelerometer. When static, the quantum accelerometer provides discrete measurements of the projection due to gravity $g_{\rm proj}$. The closed-loop scheme on the right shows how the classical accelerometer is periodically bias-corrected by comparing its output to that of the quantum accelerometer. Here, the output of the hybrid accelerometer is continuous and functions in both static and dynamic cases---providing the sum of the projections due to gravity and motion-induced acceleration $a_{\rm proj}$.}
  \label{fig:Hybridization}
\end{figure}
%--------------------------------------------------

Figure \ref{fig:Hybridization} illustrates the two hybridizaton schemes corresponding to the open-loop and closed-loop modes of the RT system. Open-loop mode provides direct measurements of the static accelerations (\eg due to gravity) by suppressing the vibration noise in the quantum accelerometer \cite{Lautier2014}. However, this mode is not optimized for operating under dynamic conditions where the QuAT is moving because the Doppler shift is not actively compensated. In closed-loop mode, the quantum accelerometer provides a measure of the classical accelerometer bias, which is then subtracted from its raw output. In this mode, the hybrid accelerometer triad can be operated in almost any orientation, as well as during dynamic motion.

%--------------------------------------------------
\begin{figure}[!t]
  \centering
  \includegraphics[width=0.80\textwidth]{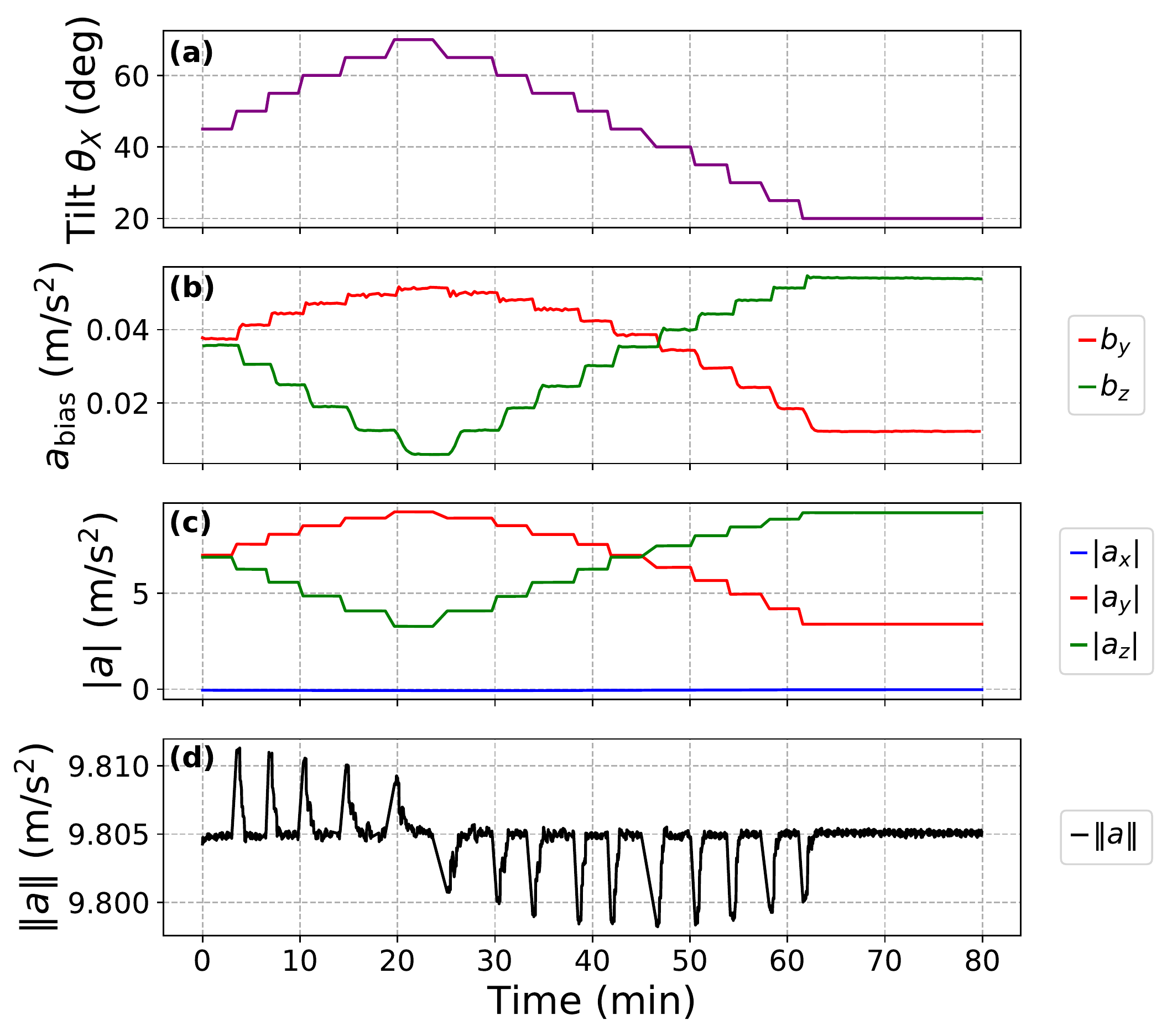}
  \caption{Output of the hybrid accelerometer triad in closed-loop mode as the tilt angle $\theta_x$ is varied in $5^{\circ}$ steps over $50^{\circ}$. Here, the interrogation time is $T = 5$ ms and $\theta_z = 0$. (a) Tilt of the triad as a function of time. (b) Classical accelerator biases measured by quantum accelerometers. These vary with the tilt angle due to a small contribution from the scale factors of the classical accelerometers. (c) Magnitude of the acceleration components. (d) Acceleration vector norm output by the hybrid triad.}
  \label{fig:HybridBias}
\end{figure}
%--------------------------------------------------

To demonstrate the capability of the closed-loop hybridization scheme, \Fig \ref{fig:HybridBias} shows acceleration measurements as the tilt angle $\theta_x$ varies in discrete steps over $50^{\circ}$. Here, the acceleration varies as $a_y = g \sin\theta_x$ and $a_z = -g \cos\theta_x$ since $\theta_z = 0$. As a result of the finite bandwidth of the central-fringe lock, the vector norm shows sharp features whenever the tilt changes abruptly. After each $5^{\circ}$ jump in $\theta_x$, the vector norm settles back to its nominal value of 9.805 m/s$^2$ regardless of the triad's orientation. Figure \ref{fig:HybridBias}(b) shows the classical accelerometer biases measured by the hybrid system. Small misalignments with the axes of the quantum triad produce coupling with the scale factors of the classical sensors---resulting in artificial bias variations with $\theta_x$.

%------------------------------------------------------------------------------------------------
\subsection*{Model of the fringe contrast}

The fringe pattern for each atom interferometer has the simple sinusoidal form:
\be
  \label{Fringe}
  P_{\mu} = A_{\mu} - B_{\mu}(\Phi_{\mu})
\ee
where $P_{\mu}$ is the probability of measuring the atom in $\ket{F = 2}$, $A_{\mu}$ is the fringe offset (typically $\simeq 1/2$), and $B_{\mu}(\Phi_{\mu})$ is an interference term given by $B_{\mu} = \frac{1}{2} C_{\mu} \cos \Phi_{\mu}$ with fringe contrast $C_{\mu}$. We showed in \Fig \ref{fig:Contrast-vs-Tilt} that $C_{\mu}$ varies strongly with the tilt of the triad due to the changing transfer efficiency of Raman transitions between the two target states. Here, we derive our model for the fringe contrast and show how it is affected by several experimental parameters. In what follows, we have dropped the subscript $\mu$ to simplify the notation.

Following the work of several other groups \cite{Hogan2009, Stoner2011, Saywell2018}, for an ideal two-level atom, we can derive an analytical expression for the fringe pattern. Under the rotating-wave approximation and ignoring spontaneous emission, the Hamiltonian describing a two-level atom interacting with a laser field is given by:
\be
  \label{Hj}
  \hat{H}(t) = \frac{\hbar}{2}
  \begin{bmatrix}
    \delta(t) & \tilde{\Omega}(t) e^{-i \phi(t)} \\
    \tilde{\Omega}(t) e^{i \phi(t)} & -\delta(t) \\
  \end{bmatrix}.
\ee
In our case, $\delta(t)$ represents the two-photon Raman detuning, $\tilde{\Omega}(t) = \sqrt{\delta^2 + \Omega^2}$ is the generalized Rabi frequency with two-photon Rabi frequency $\Omega(t)$, and $\phi(t)$ is the phase of the laser field at time $t$. The time-evolution of a wavefunction in the basis of bare-atom states: $\ket{\psi(t)} = \alpha_0(t) \ket{1,p} + \beta_1(t) \ket{2,p+\hbar k}$ can be obtained from the following unitary transformation:
\be
  \label{Unitary}
  \hat{U}(t,t') = \hat{\mathcal{T}} \exp \left[ -\frac{i}{\hbar} \int_{t}^{t'} \hat{H}(u) \dd u \right],
\ee
where $\hat{\mathcal{T}}$ is the time-ordering operator, and $\alpha_0$ and $\beta_1$ are complex state amplitudes that obey $|\alpha_0(t)|^2 + |\beta_1(t)|^2 = 1$. Henceforth, quantities labeled with the subscript $j$ represent the value of that quantity during the $j^{\rm th}$ Raman pulse with duration $\tau_j$. For short optical pulses, we can approximate the optical phase, detuning, and Rabi frequency in Hamiltonian \eqref{Hj} to be time-independent. The matrix exponential in \Eq \eqref{Unitary} then has a closed-form expression:
\be
  \hat{U}(t_j,t_j+\tau_j) \equiv \hat{U}_j = \exp \left[ -\frac{i}{\hbar} \hat{H}(t_j) \tau_j \right] = \begin{bmatrix}
    c_j^* & -i s_j^* \\
    -i s_j & c_j \\
  \end{bmatrix},
\ee
where the matrix elements contain
\be
  \label{cj_sj}
  c_j = \cos \left( \frac{\Theta_j}{2} \right) + i \frac{\delta_j}{\tilde{\Omega}_j} \sin \left( \frac{\Theta_j}{2} \right), \;\;\;\;\;
  s_j = \frac{\Omega_j e^{i\phi_j}}{\tilde{\Omega}_j} \sin \left( \frac{\Theta_j}{2} \right),
\ee
with pulse area $\Theta_j \equiv \tilde{\Omega}_j \tau_j$.

To obtain the fringe pattern $\eqref{Fringe}$ for a Mach-Zehnder atom interferometer, which consists of a $\pi/2-\pi-\pi/2$ sequence of Raman pulses, we compute the probability for the transition $\ket{1,p} \to \ket{2,p+\hbar k}$ resulting from the corresponding product of unitary transformations:
\begin{subequations}
\begin{align}
  P & = |\bra{2} \hat{U}_3 \hat{U_2} \hat{U}_1 \ket{1}|^2 \equiv A - B, \\
  A & = |s_1|^2 |s_2|^2 |s_3|^2 + |c_1|^2 |s_2|^2 |c_3|^2 + |s_1|^2 |c_2|^2 |c_3|^2 + |c_1|^2 |c_2|^2 |s_3|^2, \\
  B & = c_1 s_1 (s_2^*)^2 c_3^* s_3 + \rm{c.c.}
\end{align}
\end{subequations}
When the pulse areas $\Theta_1 = \Theta_3 = \pi/2$ and $\Theta_2 = \pi$, the offset $A$ reduces to
\be
  \label{A_Offset}
  A = \frac{\Omega_1^2 \delta_2^2 \delta_3^2 + \tilde{\Omega}_1^2 \delta_2^2 \Omega_3^2 +  \Omega_1^2 \Omega_2^2 \tilde{\Omega}_3^2 + \delta_1^2 \Omega_2^2 \tilde{\Omega}_3^2 + \delta_1^2 \Omega_2^2 \delta_3^2}{2 \tilde{\Omega}_1^2 \tilde{\Omega}_2^2 \tilde{\Omega}_3^2} \simeq \frac{1}{2}.
\ee
The last line corresponds to the near-resonant case when $|\delta_j| \ll \Omega_j$. Similarly, the interference term can be written as
\be
  \label{B_Interference}
  B = 2 |c_1| |s_1| |s_2|^2 |c_3| |s_3| \cos \Phi \equiv \frac{1}{2} C \cos \Phi,
\ee
where $\Phi = \mbox{arg}[c_1 s_1 (s_2^*)^2 c_3^* s_3] \simeq \phi_1 - 2\phi_2 + \phi_3$ is the phase shift introduced by the laser field. Here, we have omitted phase contributions from the atom's external motion relative to the laser field, as well as small terms due to $\mbox{arg}(c_1 c_3^*)$ that are linked to the asymmetry of the Mach-Zehnder interferometer \cite{Gillot2016}. Combining \Eqs \eqref{cj_sj} and \eqref{B_Interference}, and keeping only leading-order terms, the contrast is given by
\be
  \label{C_Contrast}
  C = 4 \frac{\Omega_1 \Omega_2^2 \Omega_3}{\tilde{\Omega}_1^2 \tilde{\Omega}_2^2 \tilde{\Omega}_3^2} \sin \frac{\Theta_1}{2} \sin^2 \frac{\Theta_2}{2} \sin \frac{\Theta_3}{2} \left( \delta_1 \delta_3 \sin \frac{\Theta_1}{2} \sin \frac{\Theta_3}{2} - \tilde{\Omega}_1 \tilde{\Omega}_3 \cos \frac{\Theta_1}{2} \cos \frac{\Theta_3}{2} \right).
\ee
This expression is equivalent to \Eq (3) in \Ref \cite{Gillot2016} with an explicit dependence on the two-photon detuning during each pulse. This allows us to model the behaviour of the contrast for uncompensated Doppler shifts (\ie when the RT system is disabled), as well as the effect of the velocity distribution by replacing $\delta_j = k v(t_j)$ and evaluating the velocity-averaged contrast:
\be
  \label{Cbar}
  \bar{C} = \int N(v) C(v) \dd v.
\ee
Here, $N(v) = \exp[-(v/\sigma_v)^2]/\sqrt{\pi} \sigma_v$ is the velocity probability density with $1/e$ velocity spread $\sigma_v = \sqrt{2 k_B \mathbb{T}/M}$.

Equation \eqref{C_Contrast} is valid for a two-level atom and is generally accurate when the Doppler shift $\omega_{\rm D} \gg \Omega, \omega_{\rm rec}$, where losses due to parasitic Raman transitions can be neglected. However, this condition is not satisfied when the Raman beam approaches horizontal and the effects due higher-order diffraction and velocity-insensitive transitions must be included. Using an approach similar to \Ref \cite{Hartmann2020}, we model these effects by numerically solving the full system of coupled differential equations given by:
\begin{subequations}
  \label{RamanDiffraction}
\begin{align}
  \dot{\alpha}_{n}
  & = i \chi e^{+i [\delta + 2\omega_{\rm D} + 2n\omega_{\rm rec}] t} \beta_{n-1}
    + i \chi e^{+i [\delta - 2n\omega_{\rm rec}] t} \beta_{n+1}
    + i \chi_{\rm co} e^{+i (\delta + \omega_{\rm D} + \omega_{\rm rec}) t} \beta_{n}, \\
  \dot{\beta}_{n}
  & = i \chi_{\rm co} e^{-i (\delta + \omega_{\rm D} + \omega_{\rm rec}) t} \alpha_{n}, \\
  \dot{\alpha}_{n+1}
  & = i \chi_{\rm co} e^{+i (\delta + \omega_{\rm D} + \omega_{\rm rec}) t} \beta_{n+1}, \\
  \dot{\beta}_{n+1}
  & = i \chi e^{-i [ \delta - 2n \omega_{\rm rec}] t} \alpha_{n}
    + i \chi e^{-i [ \delta + 2\omega_{\rm D} + 2(n+2) \omega_{\rm rec}] t} \alpha_{n+2}
    + i \chi_{\rm co} e^{-i (\delta + \omega_{\rm D} + \omega_{\rm rec}) t} \alpha_{n+1}.
\end{align}
\end{subequations}
Here, $\alpha_n(t)$ and $\beta_n(t)$ are the probability amplitudes corresponding to the ground and excited states $\ket{1,p+n\hbar k}$ and $\ket{2,p+n\hbar k}$, respectively; $\chi \equiv \Omega/2$ is the half-Rabi frequency for velocity-sensitive Raman transitions, $\chi_{\rm co} \equiv \Omega_{\rm co}/2$ is the half-Rabi frequency for velocity-insensitive co-propagating transitions, and $\delta$ is given by \Eq \eqref{delta}. We solve \Eqs \eqref{RamanDiffraction} over the time interval of each pulse ($t_j \to t_j + \tau_j$) to obtain the state amplitudes $\alpha_{n=0,j}$ and $\beta_{n=1,j}$. Respectively, these amplitudes are generalizations of the $c_j$ and $s_j$ when including higher-order momentum transfer. Hence, following \Eq \eqref{B_Interference}, we compute the fringe contrast using
\be
  \label{C_Contrast2}
  C = 4 |\alpha_{0,1}| |\beta_{1,1}| |\beta_{1,2}|^2 |\alpha_{0,3}| |\beta_{1,3}|.
\ee

%================================================================================================

\bibliography{References}
\bibliographystyle{ScienceAdvances}

%================================================================================================
% Acknowledgments should be gathered into a paragraph after the final numbered reference. This section should also include
% * complete funding information,
% * a description of each author's contribution to the paper,
% * a listing of any competing interests of any of the authors (all authors must also fill out the Conflict of Interest form), and,
% * a section on data and materials availability, information about the location of the data if not included in the paper, including **accession numbers** to any data relating to the paper and deposited in a public database.

\noindent \textbf{Acknowledgements:} The authors would like to thank Olivier Jolly of iXblue for his assistant developing the FPGA electronics for the real-time system. The principles behind the hybrid accelerometer triad and the real-time compensation system have been patented (US Patent No. US11175139B2, European Patent Application No. EP18709698.7A). PB acknowledges support by the Dutch National Growth Fund (NGF), as part of the Quantum Delta NL programme. The authors declare that they have no competing financial interests.

\noindent \textbf{Funding:} This work is supported by the French national agencies ANR (l'Agence Nationale pour la Recherche) and DGA (D\'{e}l\'egation G\'{e}n\'{e}rale de l'Armement) under grant no.~ANR-17-ASTR-0025-01, and ESA (European Space Agency) under grant no.~NAVISP-EL1-013. P. Bouyer thanks Conseil R\'{e}gional d'Aquitaine for the Excellence Chair.

\noindent \textbf{Author contributions:} PB, FN, B.Battelier, and B.Barrett conceived the project. ST, PC, and B.Barrett built the apparatus. ST, PC, QAC, BG, and B.Barrett performed experiments. ST and B.Barrett carried out the data analysis. All authors contributed to writing the manuscript.

%\noindent \textbf{Competing interests:} The authors declare that they have no competing financial interests.

%\noindent \textbf{Data and materials availability:} Access to raw data will be provided by the authors upon request.

%%%%%%%%%%%%%%%%%%%%%%%%%%%%%%%%%%%%%%%%%%%%%%%%%%%%%%%%%%%%%%%%%%%%%%%%%%%%%%%%%%%%%%%%%%%%%%%%%
\end{document}

%%%%%%%%%%%%%%%%%%%%%%%%%%%%%%%%%%%%%%%%%%%%%%%%%%%%%%%%%%%%%%%%%%%%%%%%%%%%%%%%%%%%%%%%%%%%%%%%%


\baselineskip24pt


\maketitle

\noindent
\textbf{This PDF file includes:}
\begin{itemize}
    \item Supplementary text
    \item Figure \ref{fig:Tracking_Lock}
    \item Table \ref{tab:RTModes}
    \item References (1 to 8)
\end{itemize}

\subsection*{Real-time phase and frequency control}
\label{sec:Realtime}

To address the issues of vibration noise and Doppler shift compensation on each axis of the interferometer, we developed a real-time (RT) phase and frequency compensation system inspired by \Ref \cite{Lautier2014}. This system is based on a Xilinx Artix-7 field-programmable gate array (FPGA) that acquires the three classical accelerometer signals, computes the appropriate frequencies and phases, and communicates them to a direct digital synthesizer (DDS) within our radio-frequency (RF) chain. This RF chain derives an ultra-low phase noise signal at 6.834 GHz that is sent to the optical IQ modulator that generates the two optical Raman frequencies. Through the DDS, we directly control the Raman phase and frequency difference at the micro-second timescale. Using high-speed serial communication with the DDS, the FPGA allows us to adjust these quantities at each Raman pulse.

For each axis of the QuAT, the FPGA performs the following operations. Between each Raman pulse (which occur at times $t_1 = t_0, t_2 = t_0 + T+2\tau_{\mu}$, and $t_3 = t_0 + 2T + 4\tau_{\mu}$, where $t_0$ is the initial time-of-flight), the RT system calculates the Doppler shift from the time integral of the acceleration $a_{\mu}^{\rm cl}(t)$ measured by the classical accelerometer:
\be
  \label{omegaRT}
  \omega_{\mu, j}^{\rm RT} = k_{\mu} \int_{t_{j-1}}^{t_j} a_{\mu}^{\rm cl}(t) \dd t
  \simeq \left\{ \begin{array}{cl}
  0 & \mbox{for pulse } j = 1, \\
  k_{\mu} (\bar{a}_{\mu} + b_{\mu}) (T + 2\tau_{\mu}) & \mbox{for pulse } j = 2, \\
  k_{\mu} (\bar{a}_{\mu} + b_{\mu}) (T + 2\tau_{\mu}) & \mbox{for pulse } j = 3. \\
  \end{array} \right.
\ee
Here, $\bar{a}_{\mu}$ is the average acceleration during the atom interferometer and $b_{\mu}$ is the bias of the classical accelerometer projected along axis $\mu$. Just before each pulse, this frequency is added to the previous DDS frequency to compensate the Doppler shift. The Raman frequency then has a step-like temporal profile given by:
\be
  \label{DeltaomegaR}
  \Delta \omega_{\mu}^{\rm R}(t) = \omega_{\rm HF} + \omega_{\rm rec} + \sum_j \omega_{\mu, j}^{\rm RT} H(t - t_j),
\ee
where $H(t)$ is the Heaviside step function. This profile can be either phase-continuous or phase-discontinuous (\ie the phase is reset to zero on each update) depending on the desired mode of the RT system, as we describe below. In addition to the frequency, the FPGA computes the following phase from the classical accelerometer signal:
\be
  \label{phiRT}
  \phi_{\mu}^{\rm RT} = k_{\mu} \int_{0}^{t_3} f_{\mu}(t) a_{\mu}^{\rm cl}(t) \dd t = k_{\mu} (\bar{a}_{\mu} + b_{\mu}) (T + 2\tau_{\mu})^2 + \phi_{\mu}^{\rm vib}.
\ee
Here, $f(t)$ is the interferometer response function \cite{Cheinet2008, Barrett2015}, and $\phi_{\mu}^{\rm vib}$ is a phase due to vibration noise that varies randomly from shot-to-shot. For practical reasons, we use a triangular-shaped response function and ignore the curvature of $f_{\mu}(t)$ during the AI pulses:
\be
  f_{\mu}(t) = \left\{ \begin{array}{cl}
    0 & \mbox{for  } t \le 0, t > 2T + 2\tau_{\mu}, \\
    t - (1 - 2/\pi)\tau_{\mu} & \mbox{for  } 0 < t \le T + 2\tau_{\mu}, \\
    2T + (3 + 2/\pi)\tau_{\mu} - t & \mbox{for  } T + 2\tau_{\mu} < t \le 2T + 4\tau_{\mu}.
  \end{array} \right.
\ee
This accounts for the pulse lengths only to first order. The phase $\phi_{\mu}^{\rm RT}$ is applied just before the final $\pi/2$-pulse to cancel the kinetic phase due to atomic motion and vibration noise due to mirror motion. We emphasize that these phase and frequency updates must be triggered before each pulse, however to avoid a phase shift that depends on the pretrigger time (20 $\mu$s in our case), the updates must occur symmetrically with respect to the temporal center of the interferometer at $t_2 = t_0 + T + 2\tau_{\mu}$ (see \Fig 11 in \Ref \citenum{Templier2021b}).

Since the real-time system effectively suppresses mirror vibrations (the primary source of phase noise in the atom interferometers), it allows us to maintain the interferometer at any location on the fringe. To scan the phase of interferometer using the laser, we simply add a controlled laser phase to \Eq \eqref{phiRT}. This also enables the use of phase modulation schemes similar to those for atomic clocks \cite{Santarelli1998} to lock the interferometer to the central fringe---maximizing the sensitivity of each quantum accelerometer.

The RT system can be configured to operate in one of two modes that we label open-loop and closed-loop. Table \ref{tab:RTModes} summarizes the resulting phase and frequency shifts for both modes. Open-loop mode operates in a similar manner to vertically-oriented atomic gravimeters \cite{Menoret2011}. Here, the raw signal from the classical accelerometer is digitally high-pass filtered on the FPGA---nullifying both the DC acceleration $\bar{a}_{\mu}$ and the accelerometer bias $b_{\mu}$. In this case, the phase computed by the RT system is simply $\phi_{\mu}^{\rm RT} = \phi_{\mu}^{\rm vib}$ and the Doppler shift compensation is implemented manually using an ``effective chirp rate'' $\alpha_{\mu}$ defined by the user. Importantly, the frequency updates are made in a phase-continuous manner such that the resulting interferometer phase is $\Phi_{\mu} \simeq (k_{\mu} \bar{a}_{\mu} - \alpha_{\mu}) (T + 2\tau_{\mu})^2$---allowing us to measure $\bar{a}_{\mu}$ from the central interference fringe.

In closed-loop mode, the digital filter is disabled and the Doppler shift is compensated as described by \Eqs \eqref{omegaRT} and \eqref{DeltaomegaR}.
In this case, the frequency steps are made in a phase-discontinuous manner to nullify their phase contribution to the total interferometer phase. Instead, the kinetic component of the interferometer is compensated by applying the phase given by \Eq \eqref{phiRT} just before the final Raman pulse. The resulting interferometer phase is $\Phi_{\mu} \simeq -k_{\mu} b_{\mu} (T + 2\tau_{\mu})^2$---providing a direct measure of the accelerometer bias $b_{\mu}$. This is the basis on which we hybridize the classical and quantum accelerometer triads, as we describe below. We note that the frequency error initially produced by the accelerometer bias is $k_{\mu} |b_{\mu}| T \simeq 2\pi \times 800$ Hz for typical bias levels of 3 m$g$. This is completely negligible compared to the linewidth of the Raman resonance ($\Omega_{\mu} \simeq 2\pi \times 50$ kHz). This error is further suppressed after a few cycles of the hybridization loop.

%--------------------------------------------------
\begin{table}[!t]
    \centering
    \begin{tabular}{ccc}
        \hline
        \hline
        Quantity & Open-loop & Closed-loop \\
        \hline
        $k_{\mu} \Delta \mu$ & $k_{\mu} \bar{a}_{\mu} (T + 2\tau_{\mu})^2 + \phi_{\mu}^{\rm vib}$ & $k_{\mu} \bar{a}_{\mu} (T + 2\tau_{\mu})^2 + \phi_{\mu}^{\rm vib}$ \\
        \hline
        $\omega_{\mu, j}^{\rm RT}$ & $\alpha_{\mu} (T + 2\tau_{\mu})$ & $k_{\mu} (\bar{a}_{\mu} + b_{\mu}) (T + 2\tau_{\mu})$ \\
        $\varphi_{\mu, j}$ & $\frac{1}{2} j (j-1) \alpha_{\mu} (T + 2\tau_{\mu})^2$ & $0$ \\
        $\phi_{\mu}^{\rm RT}$ & $\phi_{\mu}^{\rm vib}$ & $k_{\mu} (\bar{a}_{\mu} + b_{\mu})(T + 2\tau_{\mu})^2 + \phi_{\mu}^{\rm vib}$ \\
        $\Delta \phi_{\mu}$ & $\alpha_{\mu} (T + 2\tau_{\mu})^2 + \phi_{\mu}^{\rm vib}$ & $k_{\mu} (\bar{a}_{\mu} + b_{\mu})(T + 2\tau_{\mu})^2 + \phi_{\mu}^{\rm vib}$ \\
        \hline
        $\Phi_{\mu}$ & $(k_{\mu} \bar{a}_{\mu} - \alpha_{\mu}) (T + 2\tau_{\mu})^2$ & $-k_{\mu} b_{\mu} (T + 2\tau_{\mu})^2$ \\
        \hline
    \end{tabular}
    \caption{Summary of frequency and phase shifts for the two modes of the RT system. Here, $\Phi_{\mu} = k_{\mu} \Delta \mu - \Delta \phi_{\mu}$ is the total interferometer phase for axis $\mu = x, y, z$, which contains a kinetic component from the atom's motion $k_{\mu} \Delta \mu$ and a laser phase component $\Delta \phi_{\mu}$. Here, $\Delta \mu = \mu(t_1) - 2\mu(t_2) + \mu(t_3)$ is the difference between wavepacket displacements relative to the mirror during the interferometer, $\Delta \phi_{\mu} = \varphi_{\mu, 1} - 2\varphi_{\mu, 2} + \varphi_{\mu, 3} + \phi_{\mu}^{\rm RT}$ is the difference between laser phases imprinted on the atom at each pulse, with $\varphi_{\mu,j} = \int_0^{t_j}[\omega_{\mu, 2}^{\rm RT} H(t - t_2) + \omega_{\mu, 3}^{\rm RT} H(t - t_3)] \dd t$ in open-loop and $\varphi_{\mu,j} = 0$ in closed loop.}
    \label{tab:RTModes}
\end{table}
%--------------------------------------------------

%------------------------------------------------------------------------------------------------
\subsection*{Central-fringe lock}
\label{sec:FringeLock}

%--------------------------------------------------
\begin{figure}[!th]
  \centering
  \includegraphics[width=0.96\textwidth]{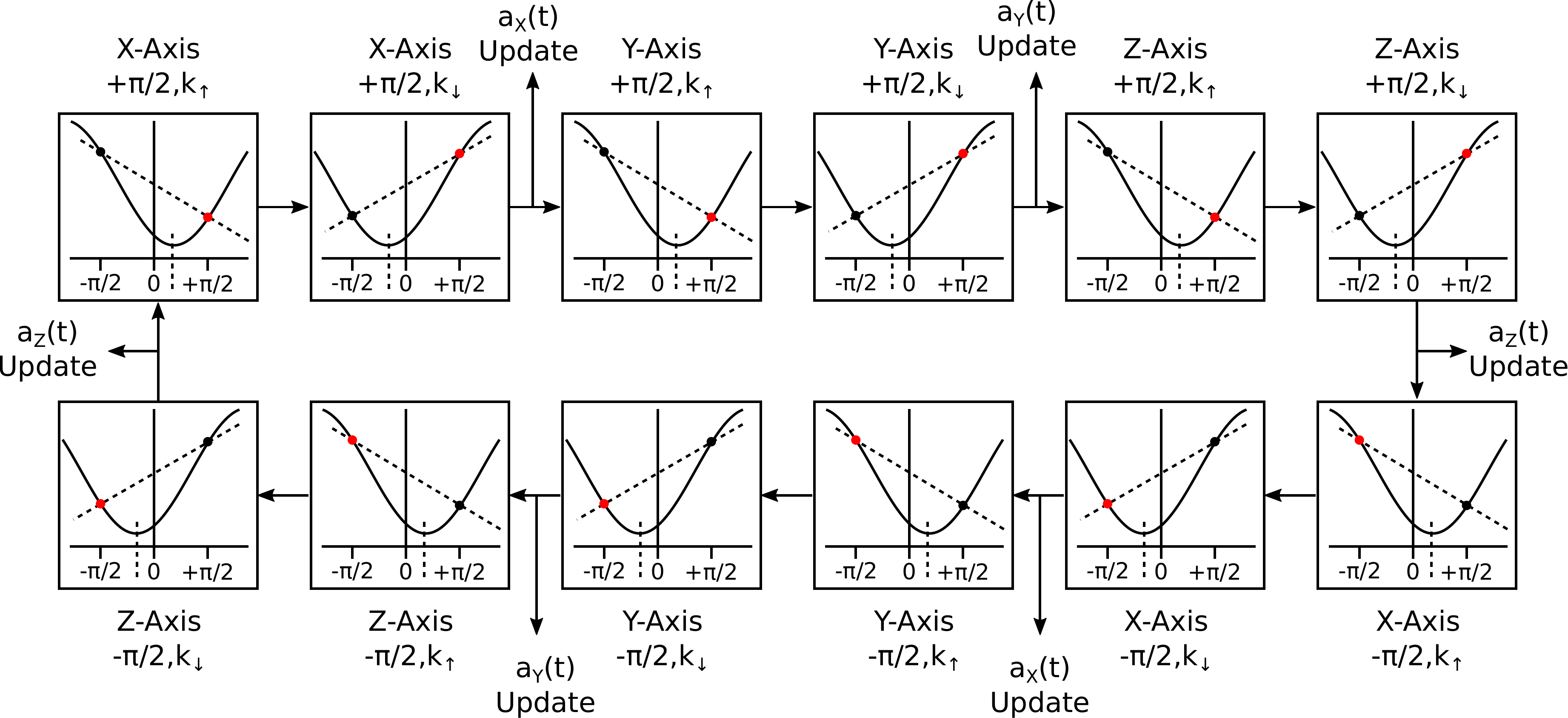}
  \caption{Central-fringe locking scheme for the QuAT. For each axis of the triad, the interferometer phase is modulated between $\pm \pi/2$ with alternating momentum transfer directions (labeled as $k_{\uparrow}$ and $k_{\downarrow}$, respectively). To initialize the feedback loop, one full cycle of 12 phase measurements is made to determine the full acceleration vector. After this, acceleration components are updated one at a time, cycling from $a_x, a_y, a_z, a_x, a_y, a_z, \ldots$, once every two measurements.}
  \label{fig:Tracking_Lock}
\end{figure}
%--------------------------------------------------

To optimize the sensitivity of the QuAT, we implemented a central fringe lock using a phase modulation protocol similar that employed by atomic clocks \cite{Santarelli1998}. As illustrated in \Fig \ref{fig:Tracking_Lock}, the phase of the interferometer is modulated between opposite mid-fringe positions ($\pm \pi/2$), where the sensitivity to phase fluctuations is largest. This scheme has the advantage of being insensitive to fringe offset and contrast variations. We also alternated between $\pm k_{\mu}$ and combined measurements such that path-independent systematic effects are rejected on the fly \cite{Merlet2009, Savoie2018}. In both open- and closed-loop mode, the total interferometer phase is steered toward zero using a simple proportional-integrator (PI) controller with an integrator time constant of $\sim 60 T_{\rm cyc} = 96$ s. In open-loop mode, this is achieved by acting on the effective chirp rate $\alpha_{\mu}$. In closed-loop mode, the lock is controlled through the bias $b_{\mu}$ fed back to the classical accelerometer stream. The effect of the central fringe lock is clear in the Allan deviation shown in \Fig 4 of the main text. The PI engages after $\sim 10$ s---initially creating a kink in the Allan deviation. For integration times greater than the time constant of the PI controller, the measurement sensitivity integrates as $1/\sqrt{\tau}$, as expected for white Gaussian phase noise.

%------------------------------------------------------------------------------------------------
%================================================================================================

\bibliography{References}
\bibliographystyle{ScienceAdvances}

%================================================================================================